\documentclass[twocolumn,aps,prd,showpacs,nofootinbib,superscriptaddress,
preprintnumbers,floatfix,10pt]{revtex4-1}

\usepackage{amsmath}
\usepackage{amsfonts}
\usepackage{amssymb}
\usepackage{epsfig}
\usepackage{graphicx}
\usepackage{color}
\usepackage{slashed}
\usepackage{epstopdf}
\usepackage{dsfont}
\usepackage{bbold}
\usepackage{mathtools}
\usepackage{array}
\usepackage{verbatim}

\newcommand{\PreserveBackslash}[1]{\let\temp=\\#1\let\\=\temp}
\newcolumntype{C}[1]{>{\PreserveBackslash\centering}p{#1}}

\usepackage{tikz}

\usetikzlibrary{positioning,decorations.pathmorphing,fit}

\usepackage{hyperref}

\newcommand{\STr}{\text{STr}}

\newcommand{\Eqref}[1]{Eq.~\eqref{#1}}

\newcommand{\Nf}{N_{\mathrm{f}}}
\newcommand{\Np}{N_{\mathrm{p}}}
\newcommand{\pat}{\partial_t}

\newcommand{\g}{\gamma} 
\newcommand{\barh}{\bar{h}}

\begin{document}

\preprint{}

\title{Self-organized criticality in a relativistic Yukawa theory with Luttinger fermions} 

\author{Holger Gies}
\email{holger.gies@uni-jena.de}
\affiliation{Theoretisch-Physikalisches Institut, 
Abbe Center of Photonics, Friedrich Schiller University Jena, Max Wien 
Platz 1, 07743 Jena, Germany}
\affiliation{Helmholtz-Institut Jena, Fröbelstieg 3, D-07743 Jena, Germany}
\affiliation{GSI Helmholtzzentrum für Schwerionenforschung, Planckstr. 1, 
64291 Darmstadt, Germany} 
\author{Marta Picciau}
\email{marta.picciau@uni-jena.de}
\affiliation{Theoretisch-Physikalisches Institut, 
Abbe Center of Photonics, Friedrich Schiller University Jena, Max Wien 
Platz 1, 07743 Jena, Germany}

\begin{abstract}
 We propose and investigate a Yukawa model featuring a dynamical scalar field
     coupled to relativistic Luttinger fermions. Using the functional
     renormalization group (RG) as well as large-$\Nf$ or perturbative
     expansions, we observe the emergence of an infrared attractive partial
     fixed point in all interactions at which all couplings become RG
     irrelevant. At the partial fixed point, the scalar mass parameter is RG
     marginal, featuring a slow logarithmic running towards the regime of
     spontaneous symmetry breaking. The long-range behavior of the model is
     characterized by mass gap formation in the scalar and the fermionic sector
     independently of the initial conditions. Most importantly, a large scale
     separation between the low-energy scales and the microscopic scales, e.g.,
     a high-energy cutoff scale, is naturally obtained for generic initial
     conditions without the need for any fine-tuning. We interpret the
     properties of our model as a relativistic version of self-organized
     criticality, a phenomenon observed in specific statistical or
     dynamical systems. This entails natural scale separation and universal
     long-range observables. We determine nonperturbative estimates for
     the latter including the scalar and fermionic mass gaps.  
\end{abstract}


\maketitle

\section{Introduction}
\label{intro}

Quantum field theories with scalar fields are prototypical for
theories with phase transitions, since the potential of the scalar field can
trigger different realizations of the ground state depending on its location and
possible symmetry-breaking features in field space. Already at zero temperature,
phase transitions can occur as a function of the parameters of the models,
specifically the parameters and couplings entailed in the scalar potential. The
underlying mechanisms find application in a wide range of fields from
statistical mechanics, condensed matter physics to elementary particle physics
and cosmology.

The parameter region near a phase transition typically features critical
phenomena and is often governed by fluctuations of the scalar fields,
potentially giving rise to features of universality \cite{ZinnJustin:1989mi}. In
relativistic four-dimensional spacetimes, approaching the critical region
generically requires a fine-tuning of parameters. This is because the mass
parameter of the scalar field represents a relevant direction of the
renormalization group (RG) with a power-counting exponent near two;  the latter
implies that microscopic parameters of a model have to be tuned with quadratic
precision to certain values in order to observe near-critical features in the
long-range physics. 

A prominent example for such properties is the Higgs sector of the standard
model of particle physics where the mass of the Higgs boson characterizing the
low-energy scale is much smaller than anticipated high-energy scales such as
those of grand-unified theories or the Planck scale of gravity.  The
correspondingly anticipated necessity of fine-tuning of parameters, specifically
the scalar mass parameter, is often considered as unnatural. Solutions to this
naturalness problem
\cite{tHooft:1979rat,Giudice:2008bi,Grinstein:2007mp,Hossenfelder:2018ikr}, or
rather explanations for the seemingly specific parameter choices, have inspired
a large number of suggestions for new underlying particle physics models. 

Standard suggestions involve, for instance, a removal of the relevant direction
by postulating an additional symmetry (e.g., scale, conformal, or supersymmetry)
or by replacing the fundamental scalar by an alternative or composite degree of
freedom that becomes effective near the electroweak scale (e.g., technicolor,
little Higgs models, etc.). 

In the present work, we explore a different option which is well known in the
context of dynamical systems: if a system has a critical point that is an
attractor of the evolution the macroscopic properties can display critical
phenomena without the need to tune microscopic parameters to specific values.
This \textit{self-organized criticality} \cite{Bak:1987xua} can, for instance,
be observed in slowly driven nonequilibrium many-body systems with nonlinear
dynamics \cite{Bak:1988zz,Manna:1991,Olami:1992,Malamud:1998,Vespignani:1998}.
Translated into relativistic quantum field theory, where the transition across
scales, i.e., RG time, plays the role of time in nonequilibrium systems, this
requires an interacting scalar  model with a slow (e.g., logarithmic) RG running
near a suitably infrared (IR) attractive fixed point. This is clearly at odds
with the strongly relevant mass parameter of a scalar model near the standard
Gaussian fixed point. 

A possible scenario for self-organized criticality in a relativistic theory has
first been sketched by Bornholdt and Wetterich in \cite{Bornholdt:1992up} where the necessary critical
point has been suggested to occur in the form of a partial RG fixed point for
the running scalar-field expectation value to be stabilized by a sufficiently
large scalar anomalous dimension. It has been argued in \cite{Bornholdt:1992up}
that this stabilization could occur for a suitable matter content.

In our work, we show that a partial infrared attractive RG fixed point occurs
naturally in Yukawa theories with scalar fields interacting with relativistic
Luttinger fermions. The latter have recently been proposed as novel particle
degrees of freedom giving rise to new asymptotically free field theories
\cite{Gies:2023cnd}; the relativistic version generalizes Luttinger fermions \cite{LuttingerPhysRev.102.1030} known from various non-relativistic condensed-matter systems, e.g,  with electronic excitations near quadratic band touching points \cite{Murakami:2004zz,Moon:2012rx,Janssen:2015xga,Janssen:2016xvc,Ray:2018gtp,Dey:2022lkx,Moser:2024dmq}. The partial fixed point occurs in our model for the Yukawa coupling as
well as for all scalar self-interactions rendering all these couplings RG
irrelevant. At the same time, the partial fixed point goes along with a large
scalar anomalous dimension which renders the scalar mass parameter RG marginal.
In the course of the RG flow, the latter exhibits a slow logarithmic drift
towards the transition from the symmetric to the broken regime. Once the
transition occurs, the partial fixed points are destroyed and the system settles
in the broken phase with a gapped spectrum. No fine-tuning is needed for a scale
separation of the low-energy observables from the microscopic scales -- in
direct analogy to self-organized criticality in dynamical systems.

We define our Yukawa model in Sec.~\ref{sec:model}. For the analysis of the
model, we utilize the functional RG as detailed in Sec.~\ref{sec:RG}. A
leading-order analysis, including large-$\Nf$ or perturbative methods is
presented in Sec.~\ref{sec:LO}. A nonperturbative analysis using a local
potential approximation of the functional RG is performed in Sec.~\ref{sec:FRG},
where also first quantitative estimates for the low-energy observables are
computed. The high-energy completion of the Yukawa model in terms of an
asymptotically free fixed point of a purely fermionic model is discussed in
Sec.~\ref{sec:HEC}. We conclude in Sec.~\ref{sec:conc}. Some technical details
are described in the appendices.  

\section{The $\g_{11}$ Yukawa model} 
\label{sec:model}

Our model is inspired by a purely fermionic model of self-interacting
relativistic Luttinger fermions first discussed and analyzed in
\cite{Gies:2024dly}. The classical Euclidean action of this model reads
\begin{equation}
    S_{\text{F}}=\int d^4x 
    \left[- \bar\psi G_{\mu\nu}\partial^\mu \partial^\nu \psi 
    -\frac{\bar g}{2} (\bar\psi \g_{11}\psi)^2\right],
    \label{eqn:S11E}
\end{equation}
where $\psi$ denotes a $d_{\g}=32$ component Luttinger spinor and $\bar\psi$ its
conjugate $\bar\psi=\psi^\dagger h$ with the spin metric $h$. For given Lorentz
indices $\mu,\nu$, $G_{\mu\nu}$ denotes a $d_\g\times d_\g$ matrix which
is tracefree in Lorentz as well as spin space and satisfies the relativistic
version of the Abrikosov algebra
\cite{Abrikosov:1974a,Janssen:2015xga,Gies:2023cnd}
\begin{equation}
    \{G_{\mu\nu}, G_{\kappa\lambda}\} = - \frac{2}{d-1} 
   g_{\mu\nu}g_{\kappa\lambda}+\frac{d}{d-1} (g_{\mu\kappa} g_{\nu\lambda} + 
   g_{\mu\lambda}g_{\nu\kappa}),
   \label{eq:AbrikosovA}
\end{equation} 
where $d$ denotes the (Euclidean) spacetime dimension. In our formulas, we keep
$d$ sometimes general for illustrative purposes, but concentrate on $d=4$
dimensional spacetime for the concrete application to studies of criticality.
The matrices $G_{\mu\nu}$ can be spanned by a suitable set of Euclidean
Dirac matrices $\gamma_A$, see Appendix \ref{sec:AppA} for further details. For
spanning the nine matrices $G_{\mu\nu}$ together with the spin metric $h$,
in total 10 Euclidean Dirac matrices are required, $\gamma_1, \dots,
\gamma_{10}$. The corresponding Dirac algebra thus contains another Dirac matrix
$\gamma_{11}$ which is used to define the interaction channel in
\Eqref{eqn:S11E}. Further scalar interaction channels are also possible
\cite{Gies:2024dly}.

By
means of a Hubbard-Stratonovich transformation, the  model of \eqref{eqn:S11E}
is fully equivalent to a description involving an auxiliary scalar field,
\begin{equation}
S_{\text{FB}}=\int d^4x 
\left[- \bar\psi G_{\mu\nu}\partial^\mu \partial^\nu \psi
- \barh \phi \bar\psi \g_{11} \psi +\frac12 \bar{m}^2 \phi^2\right],
\label{eqn:SFB11}
\end{equation}
which is obvious by using the equations of motion on the classical level or
performing the Gaussian functional $\phi$ integral on a quantum level, provided
the matching condition
\begin{equation}
\bar g = \frac{\barh^2}{\bar{m}^2} \label{eq:matching11}
\end{equation}
is fulfilled. The action \eqref{eqn:SFB11} can be the starting point of a
mean-field analysis of the fermionic model as performed in \cite{Gies:2024dly}.
In the present work, we generalize the model by considering a fully dynamical
scalar field which we consider as a fundamental quantum degree of freedom.
Focusing on the perturbatively renormalizable operators, we investigate the
$\g_{11}$ Yukawa model
\begin{align}
S=\int d^4x \bigg[&- \bar\psi G_{\mu\nu}\partial^\mu \partial^\nu \psi + \frac{1}{2} \partial_\mu \phi \partial^\mu \phi\nonumber\\
&- \barh \phi \bar\psi \g_{11} \psi +\frac12 \bar{m}^2 \phi^2 + \frac{\bar\lambda}{8} \phi^4 \bigg],
\label{eqn:Sg11Yuk}
\end{align}
where we have included a scalar kinetic term and a self-interaction with bare
coupling $\bar \lambda$. Similar to the fermionic model \eqref{eqn:S11E}, also
the $\g_{11}$ Yukawa model features a $\mathbb{Z}_2$ symmetry: the model is
invariant under a simultaneous replacement of $\psi\to-e^{i\frac{\pi}{2}
\g_{11}} \psi$, $\bar\psi \to \bar\psi e^{-i\frac{\pi}{2} \g_{11}}$, and
$\phi\to-\phi$; note that $\psi$ and $\bar\psi$ are independent variables in the
quantum theory. This $\mathbb{Z}_2$ symmetry also inhibits a bare fermionic mass
term. Here and in the following, we assume the fermion fields to occur in $\Nf$ flavors.

With both the fermionic field and the scalar field exhibiting a canonical mass
dimension of one in $d=4$, the model features the renormalizable coupling
$\bar\lambda$, the superrenormalizable Yukawa coupling $\barh$ and the scalar
mass parameter $\bar{m}$ which are a priori independent. 

It is instructive to compare the model \eqref{eqn:Sg11Yuk} with a standard
Yukawa model involving, e.g., Dirac fermions. This analogous case features the
same three parameters, all of which exert a qualitative and quantitative
influence on the low-energy observables 
\cite{Holland:2003jr,Holland:2004sd,Branchina:2005tu,Branchina:2008pc,Gies:2013fua}.
For instance, the mass parameter $\bar{m}^2$ governs the properties of the
low-energy phase: for  $\bar{m}^2$ larger than a critical value
$\bar{m}_{\text{cr}}^2$, the model remains in the symmetric phase with a massive
scalar and massless fermions and $\bar{h}$ and $\bar{\lambda}$ governing their
interactions. For $\bar{m}^2$ smaller than a critical value, the model is in the
broken phase with the scalar field acquiring an expectation value $v$
(determined by $\bar{m}^2$), all modes are gapped, and the couplings $\bar{h}$
and $\bar{\lambda}$ determine the resulting fermion mass and the mass of the
scalar $\sigma$-type mode. 

Moreover, for generic initial values of the bare mass $\bar{m}^2$, say of order
of a UV cutoff scale $\Lambda$, also the dimensionful low-energy observables,
e.g., the vacuum expectation values and the mass spectrum, will be of the order
of the cutoff scale. In order to reach a sizable scale separation with $v\lll
\Lambda$, the bare mass parameter has to be fine-tuned extremely close to the
critical value $\bar{m}_{\text{cr}}^2$. In the language of statistical physics,
the standard Yukawa model with Dirac fermions has a second-order (quantum) phase
transition with the scalar mass parameter serving as the control parameter. The
long-range physics becomes insensitive to the microscopic realization, i.e., the
cutoff can be sent to infinity, provided that the model is fine-tuned to
criticality. The latter corresponds to non-generic initial conditions, and the
fine-tuning has to be done ``by hand'' for concrete numerical solutions.  

By contrast, the model \eqref{eqn:Sg11Yuk} has very different features as we
show in the following: the model is critical for generic choices of initial
conditions, i.e., $v\lll \Lambda$ can be reached without fine-tuning, the system
is always in the broken phase, and the low-energy
observables are universal to a large degree, i.e., the mass spectrum is
independent of the bare parameters for a large region in parameter space. In
this region, the model \eqref{eqn:Sg11Yuk} has only one parameter instead of
the expected three; this one parameter essentially corresponds to a scale thus
reflecting the paradigmatic field theory property of dimensional transmutation. 

\section{Renormalization flow}
\label{sec:RG}

While the model \eqref{eqn:Sg11Yuk} can straightforwardly be analyzed using
perturbation theory or effective field theory methods, we use the functional RG
in the present work. This method has the advantage of being able to account for
threshold phenomena as they can occur in both the symmetric and the broken
regime at various scales in the present model. Perturbative or
effective-field-theory results are provided by our functional RG analysis in the
corresponding simplifying limits. 

More specifically, we employ the Wetterich equation \cite{Wetterich:1992yh}
describing the RG flow of the effective action $\Gamma_k$ as a function of an RG
scale parameter $k$,
\begin{equation}
    \pat \Gamma_k= \frac{1}{2} \STr \left[ \frac{\pat R_k} 
   {\Gamma_k^{(2)}+R_k} \right],
    \label{eq:Wetterich}
\end{equation}
where $\pat= k \frac{d}{dk}$, and $R_k$ specifies the Wilsonian momentum-shell
regularization, see
\cite{Berges:2000ew,Pawlowski:2005xe,Gies:2006wv,Delamotte:2007pf,Braun:2011pp,Dupuis:2020fhh}
for details. The bare action \eqref{eqn:Sg11Yuk} of our model serves as the
initial condition for $\Gamma_k$ at a UV scale, $\Gamma_{k=\Lambda}= S$. At
$k=0$, the action corresponds to the full quantum effective action, i.e., the
1PI generating functional $\Gamma_{k=0}=\Gamma$. 

Our approximation to solve the Wetterich equation is based on the ansatz
\begin{align}
    \Gamma_k=\int d^4x \bigg[&- Z_\psi \bar\psi G_{\mu\nu}\partial^\mu \partial^\nu \psi + \frac{Z_\phi}{2} \partial_\mu \phi \partial^\mu \phi\nonumber\\
    &- \barh \phi \bar\psi \g_{11} \psi + U(\phi) \bigg],
    \label{eqn:Gamg11Yuk}
\end{align}
where we include a full scale-dependent effective potential $U$ for the scalar
field, and also the Yukawa coupling $\barh$ and the wave function
renormalizations $Z_{\psi,\phi}$ are considered to be $k$ dependent. This ansatz
corresponds to an improved local potential approximation (so-called LPA') which
can be considered as a leading order in a systematic derivative expansion of the
action. This approximation is well tested in a plethora of nonperturbative
analyses of Yukawa systems \cite{Jungnickel:1995fp,
Hofling:2002hj,Diehl:2009ma,Braun:2010tt,Mesterhazy:2012ei,Jakovac:2014lqa,
Janssen:2014gea,Vacca:2015nta,Gies:2014xha,Classen:2015mar,Knorr:2016sfs,Fu:2016tey,Stoll:2021ori,Gies:2023jzd}.

Upon insertion into the Wetterich equation \eqref{eq:Wetterich}, our ansatz
\eqref{eqn:Gamg11Yuk} yields flow equations for all $k$-dependent quantities. It
is convenient to express the resulting flows in terms of dimensionless
renormalized quantities. For this, we first define the dimensionless effective
potential as a function of a dimensionless renormalized field invariant,
\begin{equation}
    u(\rho)= \frac{U(\phi)}{k^d}, \quad 
    \rho = \frac{Z_\phi}{2} \frac{\phi^2}{k^{d-2}}. \label{eq:uofrho}
\end{equation}
The dimensionless renormalized Yukawa coupling reads
\begin{equation}
    h^2=\frac{\barh^2}{Z_\psi^2 Z_\phi k^{6-d}}.
    \label{eq:hdimless} 
\end{equation}
The flow of the wave function renormalizations is encoded in the anomalous
dimensions
\begin{equation}
    \eta_{\psi,\phi}= - \pat \ln Z_{\psi,\phi}. \label{eq:etas}
\end{equation}
Correspondingly, the resulting flows can be written as
\begin{eqnarray}
    \pat u(\rho) & =&  -d \, u + (d-2+\eta_\phi) \rho u' \nonumber\\
    &&+ 2 v_{d}\Big[ l_{0}^{d}\left(u' + 2\rho  u''; \eta_\phi \right) \nonumber\\
    &&\qquad -\Nf d_\g  l_{0}^{(\text{L})\,d}\left(2 \rho h^2 ; \eta_\psi \right) \Big],
  \label{eq:uflow} 
\end{eqnarray}
where primes denote derivatives with respect to the invariant $\rho$. The phase
space factor $v_d^{-1}=2^{d+1}\pi^{d/2}\Gamma(d/2)$ reduces to $v_4=1/(32\pi^2)$
in $d=4$. Here and in the following, the functions $l$ (and $m$ used below)
represent threshold functions which encode the result of the regularized loop
integration. For zero arguments, they yield simple positive constants. For large
first arguments, they vanish thereby encoding threshold effects. The precise
form depends on the choice of the regulator, details are given in
App.~\ref{app:thresh} The flow of the Yukawa coupling yields
\begin{eqnarray}
    \pat h^2&=& -(2-2\eta_\psi -\eta_\phi) h^2 \nonumber\\
    &&+ 8v_d \, h_{k}^{4} 
    l_{1,1}^{\text{(LB)}\,d}\left(\omega_{1},\omega_{2};
         \eta_{\psi},\eta_{\phi}\right),  \label{eq:flowhq}
\end{eqnarray}
where 
\begin{equation}
\omega_1  =  2 \, \kappa \, h^2, \quad
\omega_2  =  u'(\kappa) + 2 \kappa u''(\kappa)
\, , \label{eq:omegadef}  
\end{equation}
and $\kappa=\rho_{\text{min}}$ denotes the minimum of the potential such that
$u'(\kappa)=0$ for $\kappa\neq 0$. The anomalous dimensions derived from the
flow of the wave function renormalizations are given by
\begin{eqnarray}
\eta_\psi\!&=& \frac{16}{d(d+2)} v_d h^2 m_{1,2}^{(\text{LB}),d}(\omega_1,\omega_2;\eta_\psi,\eta_\phi), \label{eq:etapsi}\\
\eta_\phi\!&=& \frac{8v_d}{d} \kappa (3u'' +2\kappa u''')^2 m_{2,2}^{d}(\omega_2;\eta_\phi) \nonumber\\
&&\!\!+\frac{8v_d}{d}\Nf d_\gamma h^2 \left(m_4^{(\text{L}),d}(\omega_1,\eta_\psi) -2 h^2\kappa m_2^{(\text{L}),d}(\omega_1,\eta_\psi)\! \right). \nonumber\\
&& \label{eq:etaphi}
\end{eqnarray}
Upon insertion of the solutions of Eqs.~\eqref{eq:etaphi}, \eqref{eq:etapsi}
into Eqs.~\eqref{eq:uflow}, \eqref{eq:flowhq}, the flow of the scalar potential
and of the Yukawa coupling can be integrated and the low energy observables can
be determined within the present ansatz. 

For a simplified discussion, a polynomial expansion of the potential is useful.
In the symmetric regime (SYM), we use an expansion about zero scalar field
amplitude, whereas we expand about the nontrivial minimum $\kappa>0$ in the
symmetry broken regime (SSB),
\begin{equation}
u(\rho)\simeq \left\{ 
    \begin{aligned}
        &\sum_{n=1}^{\Np} u_n \rho^n, \quad \text{(SYM)},\\
        &\sum_{n=2}^{\Np} u_n (\rho-\kappa)^n, \quad \text{(SSB)},
    \end{aligned}
\right.
\label{eq:upolexp}
\end{equation}
where $\Np$ specifies the order of the polynomial approximation as well as the
number of operators included for the parametrization of the potential. In the
simplest approximation, $\Np=2$, we use
\begin{equation}
    u(\rho)\simeq \epsilon \rho + \frac{1}{2} \lambda \rho^2\,\,\text{(SYM)}, \quad u(\rho) \simeq \frac{1}{2} \lambda (\rho-\kappa)^2\,\,\text{(SSB)},
\end{equation}
such that $\epsilon = \frac{\bar{m}^2}{Z_\phi k^2}$ denotes the dimensionless
renormalized mass, $\lambda \equiv 2 u_2 = \frac{\bar\lambda}{Z_\phi^2 k^{4-d}}$
the renormalized scalar $\phi^4$ coupling, and $\kappa=\frac{v^2}{2k^{d-2}}$ is
the dimensionless version of the expectation value of the (renormalized) field
$v=\sqrt{Z_\phi} \phi_{\text{min}}$ in the SSB regime. Once the RG flows are
computed down to low scales $k$, we can straightforwardly determine estimates
for the physical observables. For instance for flows arriving in the phase with
spontaneous symmetry breaking, we obtain the vacuum expectation value, the
scalar $\sigma$-type mass and the fermion mass from 
\begin{equation}
    v= k \sqrt{2\kappa}|_{k\to0},\,\, m_\sigma=k \sqrt{2\lambda\kappa}|_{k\to0},\,\, m_\psi = k \sqrt[4]{2\kappa h^2}|_{k\to0}
    \label{eq:observables}
\end{equation} 
where we have used $d=4$, and the fermion mass $m_\psi$ gaps the spectrum of the
Luttinger fermions along the imaginary axis in the $p^2$ plane
\cite{Gies:2024dly}. 

\section{Leading-order polynomial expansion}
\label{sec:LO}

Several aspects of the RG equations can be studied analytically. Let us first
concentrate on the flow of the relevant mass parameter $\epsilon$ and the
marginal couplings $h^2$ and $\lambda$. 

In a conventional perturbative expansion, we would focus on the deep Euclidean
region by ignoring the threshold effects, i.e., set all arguments of the
threshold functions to zero. Using the functional RG flow, it is, however,
straightforward to include the threshold phenomena. For definiteness, we perform
this first analysis in the symmetric regime, assuming $\epsilon>0$ and
$\kappa=0$, implying $\omega_1=0$, $\omega_2=\epsilon$. The corresponding
expansion of the flow equations of the relevant and marginal couplings yields
\begin{eqnarray}
    \pat \epsilon&=& -(2-\eta_\phi)\epsilon 
    - \frac{3 \left(1-\frac{\eta_\phi}{6}\right)}{32\pi^2} \frac{\lambda}{(1+\epsilon)^2} \nonumber\\
     &&+ \frac{\Nf d_\gamma \left(1-\frac{\eta_\psi}{6} \right)}{8\pi^2} h^2, \label{eq:pateps}\\
    \pat \lambda &=& 2\eta_\phi \lambda 
    + \frac{9\left(1-\frac{\eta_\phi}{6}\right)}{16\pi^2} 
    \frac{\lambda^2}{(1+\epsilon)^3} \nonumber\\
    && - \frac{\Nf d_\gamma \left(1-\frac{\eta_\psi}{6}\right)}{2\pi^2} h^4,
     \label{eq:patlam}\\
    \pat h^2 &=& - (2-2 \eta_\psi - \eta_\phi) h^2 \nonumber\\
    && + \frac{1}{8\pi^2} \frac{h^4}{(1+\epsilon)} 
    \left[ 2\left(1- \frac{\eta_\psi}{6}\right)
    +\frac{1-\frac{\eta_\phi}{6}}{1+\epsilon} \right]. \label{eq:pathq}
\end{eqnarray}
To this order, the expansion of the anomalous dimensions reads
\begin{eqnarray}
    \eta_\psi &=& \frac{\left(1-\frac{\eta_\phi}{2}\right)}{48\pi^2} 
    \frac{h^2}{(1+\epsilon)^2}, 
    \label{eq:etapsipert}\\
    \eta_\phi &=& 
    \frac{5\Nf d_\gamma \left(1-\frac{\eta_\psi}{5}\right)}{16\pi^2} h^2. 
    \label{eq:etaphipert} 
\end{eqnarray}
The one-loop result of all flows in the conventional deep Euclidean region is
obtained by setting $\epsilon=0$, i.e., ignoring threshold effects, and upon
insertion of $\eta_{\psi,\phi}$ into Eqs.~\eqref{eq:pateps}-\eqref{eq:pathq}
and a subsequent expansion to lowest-coupling order.

\subsection{Large-$\Nf$ analysis}

The preceding equations simplify in the limit of a large number of Luttinger flavors $\Nf$. For this,  we assume $\Nf
d_\gamma \gg 1$, but $\Nf d_\gamma h^2 =const$, implying that $h^2 \sim 1/(\Nf
d_\gamma) \ll 1$.  Since $d_\gamma=32$, already $\Nf=1$ turns out to
satisfy the ``large-$\Nf$'' approximation rather well. 

\begin{figure}[t]
\includegraphics[width=0.44\textwidth]{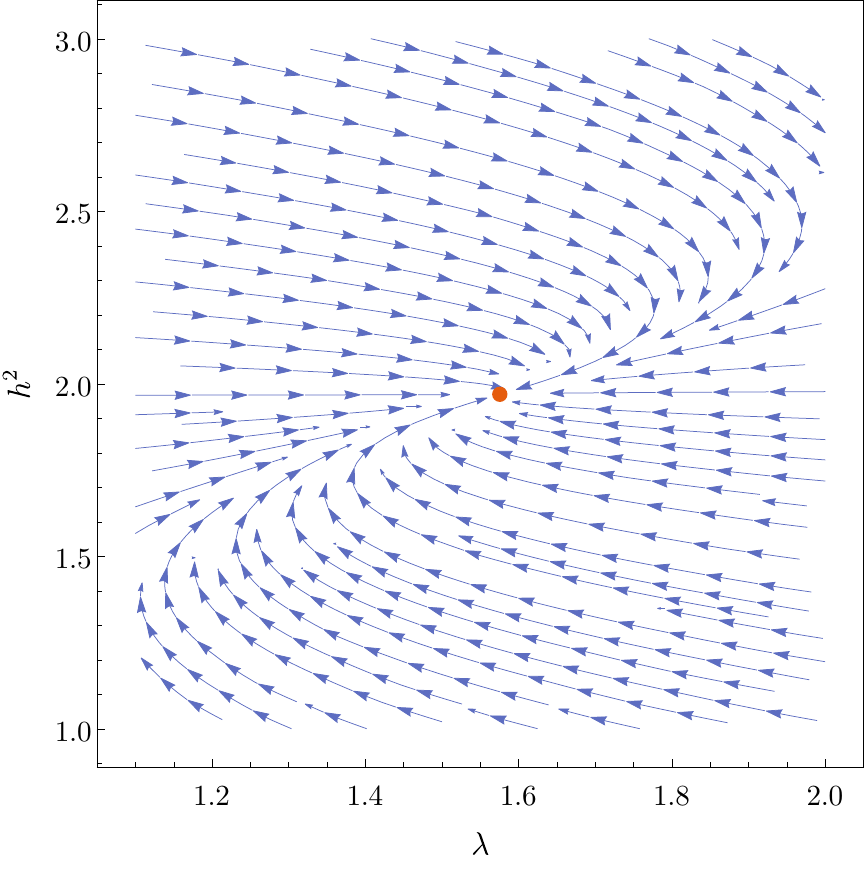}
\caption{Phase diagram of the $\g_{11}$ Yukawa model with Luttinger fermions, in
the plane of the dimensionless scalar self-interaction $\lambda$ and the
dimensionless Yukawa coupling $h^2$. The arrows indicate the RG flow towards the
IR. The interacting fixed point, highlighted in red, is fully IR attractive. The
flow has been obtained from the full equations
\eqref{eq:patlam}-\eqref{eq:pathq} with $\epsilon=10$ and for $\Nf=1$,
$d_\gamma=32$.}
\label{fig:fixedpoint}
\end{figure}

From \Eqref{eq:etapsipert}, we deduce that $\eta_\psi \simeq 0$ in this limit,
whereas $\eta_\phi$ contributes fully to leading order. Dropping the subleading
orders, the Yukawa flow \eqref{eq:pathq} reduces to
\begin{equation}
    \pat h^2 = - (2- \eta_\phi) h^2 + \mathcal{O}(1/(\Nf d_\gamma)), 
    \label{eq:pathqNf}
\end{equation}
independently of the size of $\epsilon\geq 0$. For a given value of $\Nf$ and upon insertion of $\eta_\phi\sim h^2$, the right-hand side
corresponds to a parabola in $h^2$ with two zeros. These zeros correspond to
fixed points of the RG flow. In addition to the Gaussian, i.e., non-interacting
fixed point $h^2=0$, we observe the existence of an interacting fixed point at 
\begin{equation}
h^2_\ast= \frac{32\pi^2}{5 \Nf d_\gamma} \,\, 
\Leftrightarrow\,\, \eta_\phi = 2\quad \text{for}\,\, \Nf d_\gamma\to \infty.
\label{eq:hqFPNf}
\end{equation} 
Inserting this fixed-point value into the flow of the scalar self-interaction
\eqref{eq:patlam}, also $\lambda$ exhibits a fixed point at 
\begin{equation}
 \lambda_\ast= \frac{128\pi^2}{25 \Nf d_\gamma} \quad 
 \text{for}\,\, \Nf d_\gamma\to \infty,
\label{eq:lambdaFPNf}
\end{equation} 
which demonstrates that also $\lambda_\ast$ scales like $\sim 1/(\Nf d_\gamma)$
in a large-$\Nf$ analysis. It is straightforward to check that this fixed point
is fully IR attractive in the $(\lambda,h^2)$ plane. This is illustrated in
Fig.~\ref{fig:fixedpoint} where phase diagram in terms of a stream plot of the
flow towards the IR is depicted in the $(\lambda,h^2)$ plane. 

The corresponding critical exponents, defined in terms of the negative of the
eigenvalues of the stability matrix $B$,
\begin{equation}
\theta=- \text{eig} \{B\},\quad B_{ij}= \frac{\partial (\pat g_i)}{\partial g_j}, \quad g_i=h^2,\epsilon,\lambda,\dots,
\end{equation}
are 
\begin{equation}
    \theta_{h^2}=-2, \quad \theta_{\lambda}=-4, \quad 
    \text{for}\,\, \Nf d_\gamma\to \infty,\label{eq:thetaNf}
\end{equation}
which reveals that both couplings are RG irrelevant and their flow is fully
governed by the fixed point. Since the exponents are not small, the system
approaches the fixed point rather rapidly. In conclusion, the two couplings thus
do not represent physical parameters since the long-range behavior is determined
by the fixed point. 

However, the fixed point is not a quantum scale invariant point of the full
system, but only a partial fixed point of the couplings studied so far.
Inserting the fixed-point values into the remaining equation \eqref{eq:pateps},
we find to leading order
\begin{equation}
    \pat \epsilon = \frac{4}{5}, \quad \text{for}\,\, \Nf d_\gamma\to \infty,
    \label{eq:epsFP}
\end{equation}
as a consequence of the large scalar anomalous dimension $\eta_\phi=2$. The
latter, in fact, corresponds precisely to the value that renders the mass
parameter from relevant near the Gaussian fixed point to marginal with
$\theta_m=0$ at the partial fixed point. Fixing the initial condition for
$\epsilon$ at the high scale $\Lambda$ to some value $\epsilon_\Lambda>0$, the
solution to \Eqref{eq:epsFP} reads
\begin{equation}
    \epsilon(k) = \epsilon_\Lambda + \frac{4}{5} \ln \frac{k}{\Lambda}, 
    \label{eq:epsflow}
\end{equation}
which shows that the mass parameter $\epsilon$ flows logarithmically slowly to
smaller values, transitioning into the SSB regime with
$\epsilon(k_{\text{SSB}})=0$ at
\begin{equation}
    k_{\text{SSB}}= \Lambda e^{-\frac{5}{4}\epsilon_\Lambda}.
    \label{eq:kSSBNf}
\end{equation}
We observe that $k_{\text{SSB}}\ll \Lambda$ is natural for generic choices of
$\epsilon_\Lambda$. For instance: in order to have $k_{\text{SSB}}$ being $n$
orders of magnitude smaller than $\Lambda$, we only need to choose
$\epsilon_\Lambda =n \frac{4}{5} \ln(10) \simeq 1.8 \ n $ as
initial condition. No fine-tuning is needed to separate $k_{\text{SSB}}$
from $\Lambda$. The initial conditions for $h^2$ and $\lambda$ are even
less relevant, since they are quickly attracted to the fixed point fairly
independently of the initial conditions.

Once $\epsilon(k)$ has dropped below zero for $k<k_{\text{SSB}}$,
Eqs.~\eqref{eq:pateps}-\eqref{eq:etapsipert} are no longer valid but have to be
replaced by their analogues accounting for a nontrivial minimum $\kappa$ of the
effective potential $u(\rho)$. Around $k\sim k_{\text{SSB}}$, all couplings
start to run fast. However, once the (dimensionless) minimum $\kappa$ grows
large, strong threshold effects set in since $\omega_{1,2} \sim \kappa$. As a
consequence, all threshold functions essentially drop to zero quickly describing
the decoupling of massive modes. The flow is then governed only by the
dimensional scaling terms and all dimensionful physical observables such as
those listed in \Eqref{eq:observables} approach constant values; the flow
``freezes out''. 

We emphasize that the large-$\Nf$ limit does not feature a symmetric phase,
independently of the initial conditions. Of course, large initial values for the
scalar mass term $\epsilon_\Lambda\gg 1$ will lead to a flow in the symmetric
regime over a wide range of scales, but the system will unavoidably end in the
broken phase as is obvious from \Eqref{eq:epsflow}. In the language of
statistical physics, this is reminiscent to the phenomenon of self-organized
criticality: independently of how far the seeming control parameter
$\epsilon_\Lambda$ is chosen away from a (naively anticipated) critical point,
the RG flow drives the system to criticality. The more $\epsilon_\Lambda>0$ is
chosen away from the regime transition $\epsilon=0$, the closer the system
approaches the partial fixed point in the $(h^2,\lambda)$ plane, i.e., the
closer is the system tuned to criticality with more pronounced universal
features. Nevertheless, the large-$\Nf$ flow ultimately ends up in the broken
phase with all long-range observables being naturally much smaller than the
high-energy scale $\Lambda$.  

\subsection{Perturbative analysis} 

The preceding large-$\Nf$ analysis is, in fact, more general as naively
expected, not only because $\Nf d_\gamma$ is large even for $\Nf=1$. As a
justification, let us analyse Eqs.~\eqref{eq:pateps}-\eqref{eq:etapsipert}
perturbatively without specific assumptions about the size of $\Nf d_\gamma$. In
addition to the polynomial (perturbative) dependence on the couplings $h^2$ and
$\lambda$, the equations depend non-linearly on $\epsilon$.

In the limit of large $\epsilon$, we observe that the right-hand sides of
Eqs.~\eqref{eq:pateps}-\eqref{eq:etapsipert} reproduce exactly the flow
equations of the large-$\Nf$ limit of the preceding subsection. We conclude that
the large-$\Nf$ analysis also applies to the perturbative large-$\epsilon$
regime, the latter potentially receiving  $1/\Nf$ corrections. Even the
fixed-point values $h^2_\ast$ and $\lambda_\ast$ are perturbatively small for
sufficiently large $\Nf$. 

In a straightforward naive perturbative expansion for small couplings $h^2$ and
$\lambda$, the anomalous dimensions simplify to
\begin{equation}
\eta_\psi=\frac{h^2}{48 \pi ^2 (1+\epsilon)^2} , \quad \eta_\phi
=\frac{5 h^2 \Nf d_\gamma}{16\pi^2}.
\label{eq:etaspert}
\end{equation}
Insertion into the coupling flows \eqref{eq:pateps}-\eqref{eq:pathq} and an
expansion to leading order in the couplings would yield the perturbative flows
adequately describing the vicinity of the Gaussian fixed point. 

Since we are also interested in the non-Gaussian fixed point identified before
in the large-$\Nf$ limit, there is an improved perturbative expansion which is
quantitatively more accurate also near the non-Gaussian fixed point. For this,
we observe that the non-Gaussian fixed point is characterized by a scalar
anomalous dimension $\eta_\phi\simeq 2 + \mathcal{O}(1/\Nf)$. Also the
fixed-point values for $h^2$ and $\lambda$ scale as $\sim 1/\Nf$. Inserting this
scaling into \Eqref{eq:etapsi}, we observe that $\eta_\psi$ scales like $\sim
1/\Nf^2$ near the non-Gaussian fixed point, but $\sim h^2$ near the Gaussian
one. Therefore, we can describe both fixed points by using the leading-order
formulas \eqref{eq:etaspert} to lowest perturbative order, but $\eta_\psi\sim 0
+ \mathcal{O}(1/\Nf^2)$ at higher order. Still, we keep $\eta_\phi$ as in
\eqref{eq:etaspert} also at higher order, since it appropriately accounts for
$\eta_\phi\simeq 2 + \mathcal{O}(1/\Nf)$. As a result, we obtain 
\begin{eqnarray}
    \pat \epsilon &=&-\left(2-\frac{5\Nf d_\gamma}{16\pi^2}h^2 \right)\epsilon
    -\frac{3 \lambda }{32 \pi^2 (1+ \epsilon)^2} \nonumber\\
    &&+ \frac{\Nf d_\gamma}{8\pi^2}h^2 
    \left( 1 +\frac{5\lambda}{128 \pi ^2 (1+\epsilon )^2}\right), 
    \label{eq:patepspert}\\
    \pat \lambda &=&\frac{5  \Nf  d_{\gamma }}{8 \pi ^2} h^2  \lambda 
    -\frac{ \Nf  d_{\gamma }}{2 \pi ^2} h^4 \nonumber\\
    &&  +\frac{9 \lambda ^2}{16 \pi ^2 (1+\epsilon)^3} 
    \left(1-\frac{5 \Nf d_{\gamma } }{96 \pi ^2}  h^2\right), 
    \label{eq:patlampert}\\
    \pat h^2 & = &-2 h^2 +\frac{5\Nf d_\gamma}{16\pi^2} h^4
     + \frac{5 + 3\epsilon}{12 \pi^2 (1+\epsilon)^2} h^4.  
     \label{eq:pathqpert}\\
     &&  -\frac{5\Nf  d_{\gamma }}{768\pi^4(1+\epsilon)^2}h^6 \nonumber.
\end{eqnarray}
Conventional perturbative results in the deep Euclidean region are obtained
keeping only the leading powers in $h^2$ and $\lambda$ and by setting $\epsilon
=0$ in the denominator (and in \Eqref{eq:pathqpert} also in the numerator). The
few terms that would formally be of higher order, such as the term $\sim h^2
\lambda$ in \Eqref{eq:patepspert}, the term $\sim \lambda^2 h^2$ in
\Eqref{eq:patlampert}, and the term $\sim h^6$ in \Eqref{eq:pathqpert} arise from the anomalous dimension $\eta_\phi$; they
account for the possibility that this anomalous dimension can become large
$\eta_\phi\sim \mathcal{O}(1)$ at a non-Gaussian fixed point. Nevertheless,
dropping these terms in a pure perturbative spirit would not modify the following
results qualitatively. 

From \Eqref{eq:pathqpert}, it is again obvious that the Yukawa coupling flow has
a non-Gaussian fixed point for any $\Nf$ and $\epsilon\geq 0$; for instance,
ignoring the subleading term $\sim h^6$ in \Eqref{eq:pathqpert}, the fixed-point
value assumes the compact form
\begin{equation}
h^2_\ast = \left(\frac{5 \Nf d_\gamma}{32 \pi^2} + \frac{5+3\epsilon}{24\pi^2 (1+ \epsilon)^2} \right)^{-1}. \label{eq:hqastpert}
\end{equation}
At large-$\Nf$, or alternatively large $\epsilon$, we rediscover the result of
the preceding subsection, \Eqref{eq:hqFPNf}. However, even in the extreme
opposite limit of $\epsilon=0$ and for $\Nf=1$, the numerical value for
$h^2_\ast$ deviates from the large-$\Nf$ limit only by a few percent.

The same conclusion holds for the scalar self-interaction $\lambda$ also
exhibiting a fixed-point $\lambda_\ast$ upon insertion of \Eqref{eq:hqastpert}
into \Eqref{eq:patlampert}. The somewhat extensive result can be worked out
analytically in a straightforward fashion; here we simply note that the
large-$\Nf$ result of \Eqref{eq:lambdaFPNf} is rediscovered in the corresponding
limit (and also for large $\epsilon$); in the extreme opposite limit of $\Nf=1$
and $\epsilon=0$, the result deviates only on the few percent level. 

Most importantly, this fixed point in the $(h^2,\lambda)$ plane remains fully IR
attractive for any value of $\Nf$ and $\epsilon\geq 0$. This can be read off
from the critical exponents for which we now find
\begin{equation}
    \theta_{h^2}=-2, \quad \theta_{\lambda}=[-4,-3.940\dots], \quad 
    \text{(pert.)},\label{eq:thetapert}
\end{equation}
where $\theta_\lambda$ as a function of $\Nf$ and $\epsilon$ varies in the
given interval on the few percent level; the upper end of the interval is reached for small but finite $\epsilon$. 

Also within perturbation theory, the fixed point in these couplings is, of
course, only a partial fixed point. Once it is approached rapidly in this
coupling plane, the remaining perturbative flow of the dimensionless scalar mass
parameter reads again
\begin{equation}
    \pat \epsilon = c_\epsilon, \quad \text{(pert.)},
    \label{eq:epsFPpert}
\end{equation}
where $c_\epsilon=c_\epsilon(\epsilon,\Nf)$ is a slowly varying positive function of $\epsilon$ and $\Nf$ which remains in the vicinity of its large-$\Nf$ value $c_\epsilon|_{\Nf\to\infty}=4/5$, cf. \Eqref{eq:epsFP}. E.g., for $\epsilon\to\infty$ but any $\Nf$, we find $c_\epsilon= (4/5)- 1/(20\Nf)$. In the opposite limit with $\epsilon=0$ and $\Nf=1$, $c_\epsilon$ is only about 1 percent larger.

In summary, our conclusions of the large-$\Nf$ analysis remain valid in the
whole perturbative domain: the Yukawa system develops a partial IR attractive
fixed point that is rapidly approached by the Yukawa coupling and the scalar
self-interaction for any initial value in the perturbative domain. At this
partial fixed point, the scalar mass is no longer a relevant direction, but it
is only marginal featuring a logarithmically slow running towards the regime of
chiral symmetry breaking. Again, we conclude that a large scale separation
$k_{\text{SSB}}\ll \Lambda$ is natural for generic choices of the initial
conditions.

\section{Functional RG flow}
\label{sec:FRG}

Let us now integrate the functional flows \Eqref{eq:uflow}-\eqref{eq:etaphi}
without any assumption on the size of $\Nf$ or the values of the couplings.
While there are powerful methods available to solve also the potential flow
\eqref{eq:uflow} as a partial differential equation in field space
\cite{Bervillier:2007rc,Borchardt:2015rxa,Borchardt:2016pif,%
Borchardt:2016xju,Grossi:2019urj,Koenigstein:2021syz,Ihssen:2022xkr,%
Ihssen:2023qaq,Sattler:2024ozv},
we use a simple polynomial expansion about the minimum as parametrized in
\Eqref{eq:upolexp}. This gives us access to the spectral information of the
Yukawa system, and we can monitor the convergence of this expansion as a
function of the polynomial order $N_{\text{p}}$. 

At the initial scale $k=\Lambda$, we impose nontrivial initial conditions on all
perturbatively marginal or relevant couplings necessary in order to render
the theory fully interacting, i.e., choose initial values for
$\epsilon_\Lambda$, $h^2_\Lambda>0$; for simplicity, we set all other
$u_{n\geq 2}=0$ (including $\lambda_\Lambda$) at $k=\Lambda$, as these couplings
are generated by the flow anyway.  However, since all scalar couplings are
quickly attracted by their corresponding partial fixed point with a large
negative (RG irrelevant) critical exponent, generic nontrivial initial
conditions for all other $u_{n\geq 2}$ do not take any relevant influence
on the results. 

As for the initial conditions for $\epsilon_\Lambda$ and $h^2_\Lambda$, there
are qualitatively two resulting flows: for negative
$\epsilon_\Lambda<0$ (or small $\epsilon_\Lambda>0$ with sufficiently large
$h^2_\Lambda$), the flow starts in (or runs comparatively quickly into) the
broken regime where all modes become massive and decouple quickly. In this case,
$k_\text{SSB}\lesssim \Lambda$ remains fairly close to the high scale. The
resulting dimensionful quantities such as the vacuum expectation value $v$ or
the particle masses $m_\sigma$ or $m_\psi$ depend strongly on the details of the
initial conditions. In this case, the RG flow is not governed by a (partial)
fixed point, hence we do not observe nor expect universal features.

By contrast, for sufficiently large mass parameter, say $\epsilon_\Lambda
\gtrsim O(1)$, and perturbative or medium large initial Yukawa couplings
$h^2_\Lambda$, the RG flow of all other couplings is attracted by the partial
fixed point present in all couplings $h$, $\lambda=u_2$, and all other $u_n$,
while $\epsilon$ runs logarithmically slowly towards zero and then into the SSB
regime, such that $k_{\text{SSB}}\ll \Lambda$. No fine-tuning of any of the
parameters is needed for this generic situation; in fact, the deeper we put the
system into the symmetric regime, e.g., with a large positive
$\epsilon_\Lambda$, the more RG time the systems spends near the partial fixed
point and is ultimately driven to criticality. As the partial fixed point
renders all other couplings RG irrelevant, the IR observables show a large
amount of universality, and we can express all dimensionful quantities in units
of a single scale. 

\begin{figure}[t]
    \includegraphics[width=0.48\textwidth]{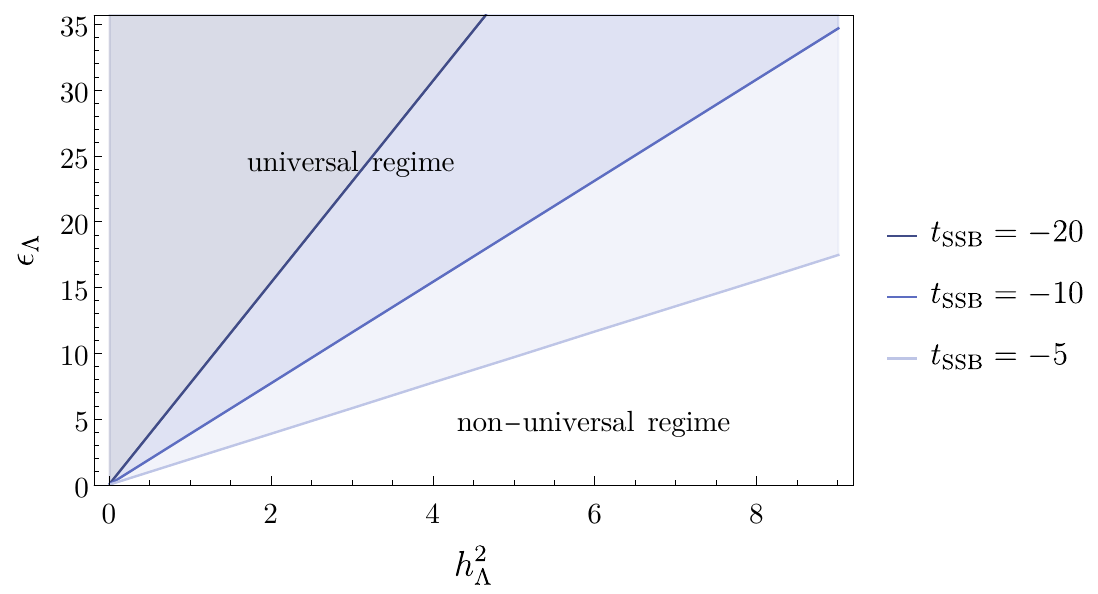}
    \caption{Degree of universality in the $(h_\Lambda^2,\epsilon_{\Lambda})$
    plane of initial conditions for the $\Nf=1$ $(d_\gamma=32)$ model. The solid
    curves mark initial conditions for which $t_{\text{SSB}}=-5,-10,-20$. The
    shaded regions above these curves correspond to generic initial conditions
    with a correspondingly increasing degree of universality. (The initial
    values of all the irrelevant couplings $u_{n\geq 2}$ are set to zero, but do
    not exert a significant influence on the data anyway, see text). In
    subsequent studies, we use $t_\text{SSB}\leq -10$ exibiting a degree of
    universality sufficient for all practical purposes.}
\label{fig:universality}
\end{figure}

The degree of universality of the long-range observables is governed by
the RG time spent at the partial fixed point; for a simple estimate, we use the
RG time scale $t_{\text{SSB}}= \ln \frac{k_{\text{SSB}}}{\Lambda}$ where the
system runs into the broken regime as a proxy for the time spent near the fixed
point. Since the irrelevant perturbations near the fixed point die out with
their corresponding critical exponents, the non-universal corrections contribute
at most with the largest exponent $\theta_{h^2}=-2$, cf. \Eqref{eq:thetaNf},
such that corrections to universality scale maximally with $\sim
(k_{\text{SSB}}/{\Lambda})^2$. Therefore, $2 (\log_{10} e) |t_{\text{SSB}}|
\simeq 0.82 |t_{\text{SSB}}|$ serves as an estimate for the number of digits of
a long-range observable that are unaffected by non-universal corrections. In
Fig.~\ref{fig:universality}, we depict the curves in the
$(h_\Lambda^2,\epsilon_{\Lambda})$ plane of initial conditions for which we
obtain $t_{\text{SSB}}=-5,-10,-20$ for the $\Nf=1$ $(d_\gamma=32)$ model. The
shaded regions above these curves exhibit universality of the long-range
observables at least to this estimated degree. Also, we haven't found any
significant influence of the initial condition for the scalar interaction
$\lambda_\Lambda$ on these curves. This is in agreement with the even more
subleading critical exponent $\theta_\lambda \simeq -4$ which induces a rapid
die out of the scalar self-coupling. In conclusion, a large region in the space
of initial parameters leads to universal long-range physics. This justifies to
call these initial conditions \textit{generic}. No fine-tuning is needed at all
to put the system into this region.

As we are mainly interested in this universal regime which we interpret as the
analogue of self-organized criticality of dynamical systems, we initialize the
flow such that  the system spends sufficient RG time $t=\ln\frac{k}{\Lambda}$ in
the symmetric regime, in order for the couplings to be sufficiently attracted by
the fixed point, before entering the broken regime. The preceding
considerations have also been confirmed by fully numerical tests demonstrating
that $t_\text{SSB}\lesssim-10$, i.e. $k_\text{SSB}/\Lambda\lesssim 10^{-5}$, is
sufficient to suppress non-universal corrections within our numerical
accuracy. For concrete computations, we set $h^2_\Lambda =1$, $
u_{n,\Lambda}=0$ and choose $\epsilon_\Lambda$ such that the universal regime
will always be reached. For the cases $\Nf=1,2$, $\epsilon_\Lambda$ has been set
to $10$. For larger $\Nf$, the initial condition $\epsilon_\Lambda$ has been
chosen somewhat larger such that the transition time $t_\text{SSB}\lesssim-10$.
This is, because larger $\Nf$ for $h_\Lambda^2$ fixed correspond to large
initial values for $\eta_\phi$, cf. for instance \Eqref{eq:etaspert}, inducing a
faster initial running until the couplings have sufficiently approached their
fixed point values.

\begin{figure}[t]
    \includegraphics[width=0.48\textwidth]{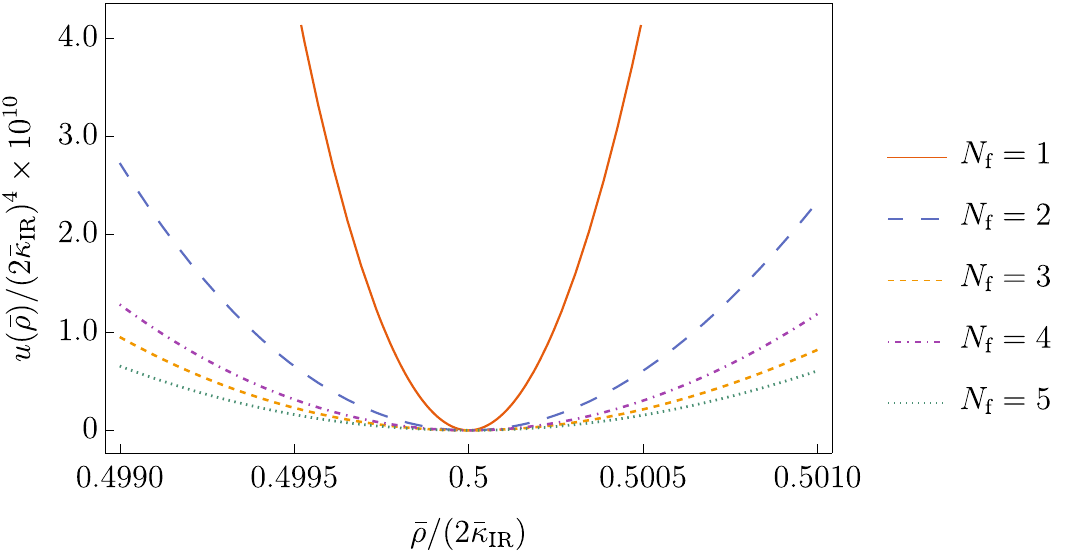}
    \caption{Dimensionful effective potential of our Yukawa model with Luttinger fermions as a function of the field amplitude $\rho=\frac{1}{2} \phi^2$ for different values of $\Nf$ and $d_\g=32$ in units of the resulting vacuum expectation value $v$. A polynomial expansion at $18^{\text{th}}$ order of the potential has been used, and initial conditions at the UV scale $\Lambda$ have been chosen such that the flow spends a sufficently wide range of scales near the partial fixed point; no fine-tuning is needed for this while the resulting potential is universal, i.e., essentially independent of the microscopic initial conditions. The figure displays the limited range of field values where the polynomial expansion passes D'Alembert's ratio test for convergence. }
\label{fig:potential}
\end{figure}

In Fig.~\ref{fig:potential}, we plot the resulting dimensionful effective
potential as a function of the dimensionful field invariant both in units of the
vacuum expectation value $v$ for various values of $\Nf$. In all cases, the
potential develops a nontrivial expectation value. Using d'Alembert's ratio
test, we have performed an estimate for the convergence radius of the polynomial
expansion. Going up to $18^{\text{th}}$  order in the expansion, the ratio test
suggests that the convergence radius is  of the order $0.005$ (in units of $v$);
the corresponding highest-order results are shown in Fig.~\ref{fig:potential}.
(NB: The polynomial expansion is not able to resolve the convexity property of
the effective potential. The full flow of \Eqref{eq:uflow} would lead to a
convex potential, implying that the potential to the left of the minimum in
Fig.~\ref{fig:potential} would become flat in the limit $k\to0$
\cite{ORaifeartaigh:1986hi,Litim:2006nn,Ihssen:2023qaq,Zorbach:2024zjx,Zorbach:2024rre}.)  

\begin{figure}[t]
    \includegraphics[width=0.48\textwidth]{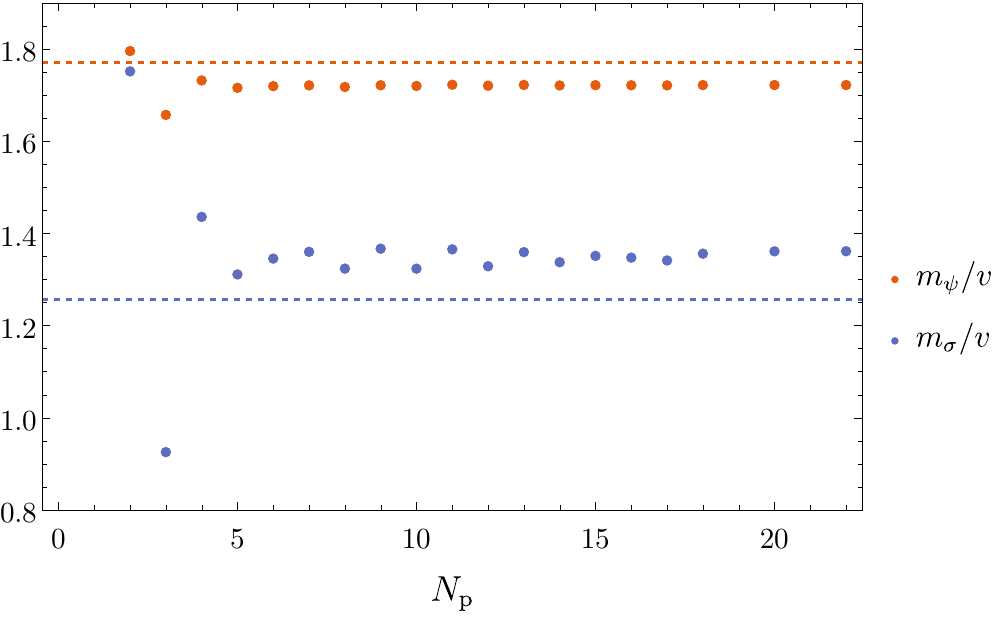}
    \caption{Ratio of the scalar $\sigma$-type mass $m_\sigma$ and the vacuum expectation value $v$ (orange) and ratio of the fermionic mass gap $m_\psi$ and $v$ (blue) for the $\Nf=1$ model ($d_\g=32$) as a function of the polynomial expansion $N_{\text{p}}$. The initial conditions have been chosen such that the model is in the universal regime. The dashed lines (in the corresponding colors) correspond to an estimate based on the fixed point analysis in the large $\Nf$ limit, c.f. \Eqref{eq:hqFPNf} and \Eqref{eq:lambdaFPNf}, as explained in the main text.}
    \label{fig:massratios}
\end{figure}

On the basis of this numerical control of the full flow of the effective
potential near its minimum, we can straightforwardly determine the mass of the
$\sigma$-like scalar excitation $m_\sigma$ as well as the fermionic mass gap
$m_\psi$ according to \Eqref{eq:observables}. In the universal regime, their
scale is clearly set by the vacuum expectation value $v$ as well. Since the RG
flow of the coupings $\lambda$ and $h^2$ is governed by the partial fixed point
for a wide range of scales, the partial fixed point for these couplings also
exerts an influence on the final mass values. Once, the scalar mass parameter
$\epsilon$ crosses zero at $k_{\text{SSB}}$, the couplings depart from their
fixed-point values such that the details of the SSB flow ultimately determine
the mass spectrum quantitatively. For $\Nf=1$, the resulting values for the mass
spectrum are shown in Fig.~\ref{fig:massratios} as a function of the
approximation order $N_{\text{p}}$ in units of the vacuum expectation value $v$.
While small values of $N_{\text{p}}$ exhibit somewhat larger truncation artefacts, the convergence with increasing order of the truncation appears satisfactory;
in particular for the highest orders $N_\text{p}=20,22$, the variation is
on the sub-permille level.
Quantitatively, we find $m_\sigma/v\simeq 1.36$ for the sigma-like mass of
the scalar excitation and $m_\psi/v\simeq 1.72$ for the fermionic mass gap.

The dashed lines indicate the would-be value of the masses if computed from the
partial fixed point values $h_\ast^2$, $\lambda_\ast$ of Eqs.~\eqref{eq:hqFPNf},
\eqref{eq:lambdaFPNf} in the large-$\Nf$ limit. More precisely, the
estimate for $m_\sigma/v$ indicated by the blue dashed line corresponds to
$m_\sigma/v\simeq \sqrt{\lambda_\ast}=\frac{2\pi}{5}$, using
\Eqref{eq:observables} and the large-$\Nf$ limit fixed-point value
\Eqref{eq:lambdaFPNf}; a similar estimate for $m_\psi/v$ involves to choose a
scale, since the Yukawa coupling in the original action is dimensionful. The
relevant scale here is $k_{\text{SSB}}$, since this is the scale where the
system starts departing from the partial fixed point and subsequently decouples.
Hence, the red dashed line in Fig.~\ref{fig:massratios} is given by the estimate
$m_{\psi}/v \simeq \frac{\sqrt[4]{h_\ast^2 k_{\text{SSB}}^2}}{\sqrt{v}}$,
where we use the large-$\Nf$ limit fixed-point value \Eqref{eq:hqFPNf} in
addition to the numerical data for $k_{\text{SSB}}$ and $v$. Since the
deviations from finite-$\Nf$ corrections are on the few percent level, cf.
\Eqref{eq:hqastpert}, the visible difference of the full numerical result (dots)
from the estimates (dashed lines) in Fig.~\ref{fig:massratios} is a result of
the full RG flow in the threshold regime $\epsilon\lesssim 0$. The fact that
this difference is only on the $O(10\%)$ level justifies the interpretation that
the properties of the long-range observables are essentially governed by the
properties of the partial fixed point. Even though the fixed point is destroyed
in the course of the transition to the SSB regime, the hierarchy of the
couplings is essentially preserved in the course of the flow through the
threshold regime.

\begin{figure}[t]
    \includegraphics[width=0.48\textwidth]{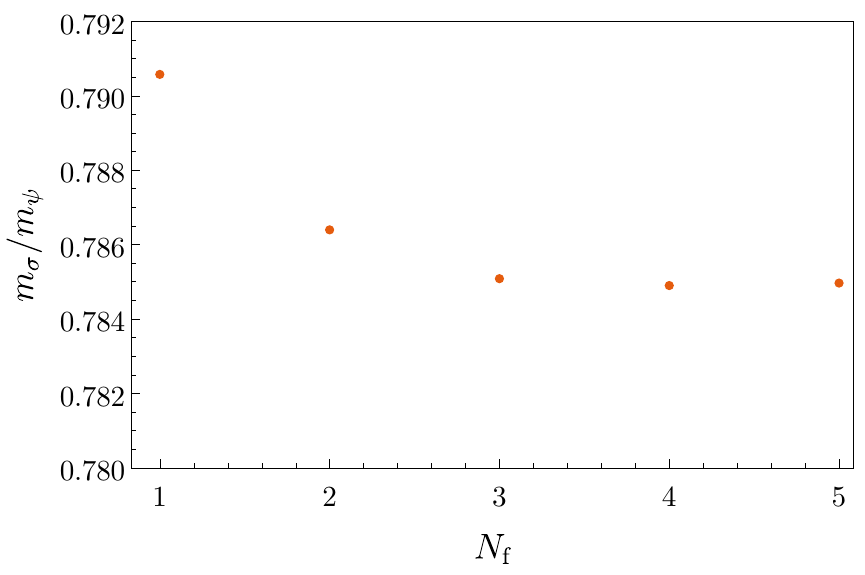}
    \caption{Ratio between the scalar $\sigma$-type mass  $m_\sigma$  and the
    fermion mass  $m_\psi$ as a function of the flavor number $\Nf$ ($d_\g=32$).
    All data points have been produced within the $N_{\text{p}}=22$
    truncation inside the universal regime.}
    \label{fig:massratiosNf1to5}
    \end{figure}

The mass ratio $m_\sigma/m_\psi$ is a particularly relevant prediction of our
model for several reasons: from the viewpoint of the high-energy completion of
the model discussed Sec.~\ref{sec:HEC} below, the scalar could arise as
bi-fermionic bound state. In this context, the deviation of the ratio from
$m_\sigma \simeq 2 m_\psi$ is a measure for the binding energy of the scalar
state. Also, for a comparison with other nonperturbative methods, we expect the
mass ratio to play a useful role; e.g., lattice simulations typically have a
direct access to spectral information via the study of spatial correlation
functions. Our result for the mass ratio is shown in
Fig.~\ref{fig:massratiosNf1to5} as a function of the flavor number $\Nf$ and for
the highest truncation order $N_{\text{p}}=22$. As expected, we observe a
variation on the percent level for small $\Nf$, rapidly converging for larger
$\Nf$. The mass ratio shown in the plot for $\Nf=5$ agrees already on the
per-mille level with $m_\sigma/m_\psi\simeq 0.786$ computed for $\Nf=100$ as a
large-$\Nf$
reference value.

From the viewpoint of the high-energy completion of the model where the scalar
is a bi-fermionic bound state, we conclude that the mass ratio near
$m_\sigma/m_\psi\simeq 0.79 < 2$ points to a deeply bound relativistic state
where the binding energy exceeds the mass gap of a single fermionic constituent.

\begin{figure}[t]
\includegraphics[width=0.48\textwidth]{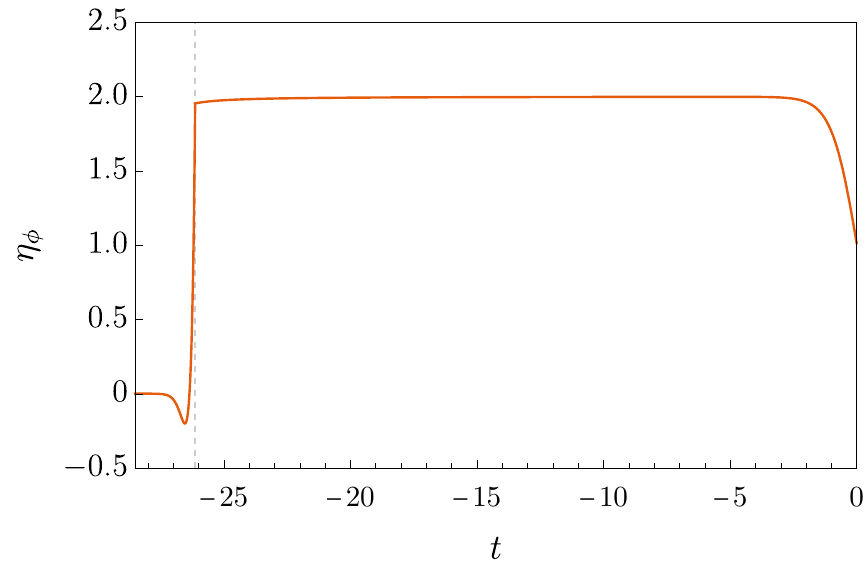}
\caption{Scalar anomalous dimension $\eta_\phi$ for a typical RG flow initiated in the universal parameter space. The plot should be read from right to left (UV to IR): starting in the symmetric regime at $t=0$ ($k=\Lambda$), $\eta_\phi$ rapidly approaches the fixed-point value $\eta_\phi=2$, c.f. \Eqref{eq:hqFPNf}, and remains there for a wide range of scales. After entering the broken phase towards small $t$, $\eta_\phi$ vanishes due to threshold effects. This plot uses $\Nf=1$, $d_\gamma=32$, $N_{\text{p}}=18$, $\epsilon_\Lambda=10$, $h_\Lambda^2=1$, $u_{n\geq 2}=0$, yielding a transition scale of  $t_{\text{SSB}}\approx-26$, i.e., $k_{\text{SSB}}\simeq 5\times 10^{-12} \Lambda$, as indicated by the gray dashed line. }
\label{fig:etaphi}
\end{figure}

Finally, the property of self-organized criticality can also be read off from
the flow of the scalar anomalous dimension. A typical flow is depicted in
Fig.~\ref{fig:etaphi} for initial conditions in the universal regime
($\epsilon_\Lambda=10$, $h_\Lambda^2=1$, $u_{n\geq 2}=0$) and the case $\Nf=1$.
Near the cutoff $k\lesssim\Lambda$ ($t\lesssim 0$), the flow rapidly approaches
the fixed-point value $\eta_\phi \simeq 2$ and remains there for a wide range of
scales. This goes hand in hand with the fact that the scalar mass parameter no
longer is a relevant operator but becomes marginal at the partial fixed point
where it runs logarithmically slowly towards the broken regime quantitatively
similar to the large-$\Nf$ flow \eqref{eq:epsFP}. At the same time,
$\eta_\phi\simeq 2$ induces the partial fixed points in all other couplings
$h^2, \lambda, u_{n>2}$ while keeping $\eta_\psi$ numerically small as expected
from \Eqref{eq:etapsipert}. This fixed-point controlled flow stops, once
$\epsilon$ drops below zero, which in the case of Fig.~\ref{fig:etaphi} happens
at exponentially small scales  near $t_{\text{SSB}}\approx -26$, i.e.,
$k_{\text{SSB}}\simeq 5\times 10^{-12} \Lambda$. Here, $\eta_\phi$ starts to run
fast towards zero. Once the RG scale drops below the scale of the vacuum
expectation value, all modes become massive and decouple which implies that
$\eta_\phi\to 0$ for $t<t_{\text{SSB}}$.

Comparing the results for our model with the scenario for self-organized
criticality in chiral Higgs-Yukawa models suggested by Bornholdt and Wetterich
in \cite{Bornholdt:1992up}, the scalar anomalous dimension $\eta_\phi$ plays the
role of the mass anomalous dimension $\omega$ defined in
\cite{Bornholdt:1992up}. The quantitative criterion for self-organized
criticality suggested  in \cite{Bornholdt:1992up} (Bornholdt-Wetterich
criterion),
\begin{equation}
 \langle \omega \rangle = \frac{1}{t_0} \int_0^{t_0} dt \ \omega(t)\simeq 2, \quad t_0 := \ln \frac{v}{\Lambda},
 \label{eq:meanomega}
\end{equation}
is evidently satisfied for $\omega=\eta_\phi$, as the scalar anomalous dimension
is essentially constant in $t$ and governed by the partial fixed point value
$\eta_\phi\simeq 2$.  While \Eqref{eq:meanomega} could be satisfied also by a
varying function $\omega(t)$, our Yukawa model based on Luttinger fermions
satisfies the Bornholdt-Wetterich criterion in a straightforward fashion. In
contrast to the scenario developed in \cite{Bornholdt:1992up}, our model bridges
the wide ranges of scales between $k=\Lambda$ and $k\simeq v$ fully in the
symmetric regime. A flow in the broken regime with $\kappa>0$ or an attractive
partial fixed point with $\kappa \to \kappa_\ast$ as studied in
\cite{Bornholdt:1992up} is not needed for self-organized criticality as featured
by our model.

\section{High-energy completion of the model}
\label{sec:HEC}

So far, we have studied the flow towards the IR, assuming that the microscopic
parameters of the model have been fixed at some initial high-energy scale
$\Lambda$. The Yukawa model exhibits a remarkable degree of universality as its
long-range physics is governed by a partial IR fixed point which is fully IR
attractive apart from the marginal mass-parameter direction.

Let us now concentrate on the high-energy behavior of the model by addressing
the question as to whether RG trajectories exist along which we can take the
limit $\Lambda\to \infty$. If so, the corresponding model is UV complete and
could exist on all length scales.

From the properties of the IR fixed point, we can already conclude that its UV
flow is fully repulsive in the Yukawa and all scalar self-couplings. Therefore,
the only RG trajectories for which we may have full UV control are those that
emanate from the partial fixed point. Other options would require the existence
of further UV fixed points; however, we haven't found any other fixed point in
the validity regime of our approximation except for the Gaussian one which would
yield a trivial free theory. The fact that all Yukawa and scalar self-couplings
are irrelevant at the partial fixed point implies that they are also irrelevant
for UV-complete flows emanating from the fixed point. The only physical
parameter to be fixed is the mass parameter. We know from the preceding studies
that the mass direction runs logarithmically if the other couplings are at the
partial fixed point. Towards the UV, the mass parameter $\epsilon$ runs
logarithmically to $+\infty$ for $k\to \infty$.

It is useful to study the flow of the ratio
\begin{equation}
g= \frac{h^2}{\epsilon}, \label{matching11b}
\end{equation}
which corresponds to the renormalized version of the matching condition of the
partially bosonized purely fermionic $\gamma_{11}$ model (with nondynamical
scalars and  zero scalar self coupling) \cite{Gies:2024dly}. We can 
straightforwardly derive the flow of this ratio in the present Yukawa model from
Eqs.~\eqref{eq:pateps} and \eqref{eq:pathq}, yielding
\begin{eqnarray}
\pat g &=& 2 \eta_\psi g + \frac{3}{32 \pi^2} 
\left(1-\frac{\eta_\phi}{6} \right)  \frac{\lambda}{\epsilon(1+\epsilon)^2} 
g\nonumber\\
&& - \frac{1}{8\pi^2} \left[ \Nf d_\gamma \left(1-\frac{\eta_\psi}{6}\right) 
\right. \nonumber\\
&& \qquad \left. - \frac{\epsilon}{1+\epsilon} \left(\left(2 - \frac{\eta_\psi}{3}\right) 
+ \frac{1- \frac{\eta_\phi}{6}}{1+\epsilon}\right) \right] g^2 .
\label{eq:patratio}
\end{eqnarray} 
In order to understand the high-energy behavior, we note that $\epsilon$ grows
large, implying $\eta_\psi\to 0$, while the coupling $\lambda$ approaches the
fixed point and thus remains bounded, as does $\eta_\phi$.  In the limit
$\epsilon \to \infty$, we obtain
\begin{equation}
\pat g = - \frac{1}{8\pi^2} (\Nf d_\gamma -2 ) g^2
\label{eq:patg}
\end{equation}
which corresponds precisely to the flow of the fermionic self-coupling in the
$\g_{11}$ model including the anticipated $1/\Nf$ corrections
\cite{Gies:2023cnd,Gies:2024dly}.

Therefore, we can interpret the UV complete RG trajectories present in our
Yukawa model as follows: the flow of the Yukawa coupling dominated by the
partial fixed point and the logarithmic running of the scalar mass parameter are
reflections of the asymptotic freedom of the purely fermionic $\gamma_{11}$
model. The latter is UV complete, features dimensional transmutation, and
exhibits the same long-range behavior as our Yukwawa model. We conclude that the
UV complete trajectories in our Yukawa model and the fermionic $\gamma_{11}$
model are in the same universality class, since they are governed by the same
fixed point. As our preceding discussion has demonstrated, the models are also
in the same universality class even if the Yukawa model is initiated with
generic initial conditions. This is because the partial fixed point is fully IR
attractive in those couplings that would induce deviations from the purely
fermionic description. Of course, if the Yukawa model serves as an effective
field theory fixed at an initial UV scale $\Lambda$, the long-range physics can
deviate from the universal trajectory by corrections of the order $\sim
1/\Lambda^{|\theta_i|}$, where $\theta_i$ corresponds to a suitable exponent of
the irrelevant operators.

This interpretation is also corroborated by the scalar anomalous dimension being
$\eta_\phi\simeq 2$ near the partial fixed point. This implies that the scalar
wave function renormalization behaves like
\begin{equation}
Z_\phi(k) \simeq Z_\phi(k_{\text{IR}}) \frac{k_{\text{IR}}^2}{k^2},\quad 
\text{for}\,\, k\to\infty
\label{eq:Zphi}
\end{equation}
for $k_{\text{IR}}$ representing some IR scale at which the field amplitude is
renormalized. For instance, normalizing the wave function renormalization
naturally to $Z_\phi(k_{\text{IR}})=1$ in the long-range limit, the wave
function renormalization becomes small towards the UV. The kinetic term of the
scalars thus becomes suppressed and the scalar field more and more resembles a
nondynamical auxiliary field, similarly to that introduced by a
Hubbard-Stratonovich transformation of a fermionic self-interaction. Of course,
near the transition scale $k\simeq k_{\text{SSB}}$, $\eta_\phi$ deviates from
$\eta_\phi\simeq 2$ and approaches $\eta_\phi\simeq 0$ for $k\ll
k_{\text{SSB}}$, implying that \Eqref{eq:Zphi} receives some quantitative
corrections; however, the scaling with $\sim 1/k^2$ towards higher energies
holds true over the range of scales where the system is close to the partial
fixed point.

Finally, the present fermionic picture also offers an explanation for the
fact that the curves of constant $t_{\text{SSB}}$ in Fig.~\ref{fig:universality}
have almost constant slope: in the fermionic language, $t_{\text{SSB}}$
essentially corresponds to the scale of the IR divergence of the coupling $g$
when integrating the flow  \eqref{eq:patg} towards the IR. Now, changing the
initial conditions for $\epsilon_\Lambda$ and $h_\Lambda^2$ such that their
ratio $g_\Lambda= h_{\Lambda}^2/\epsilon_\Lambda$ remains fixed, leaves the
scale for the IR divergence of $g$ unchanged. The universal region in the full
Yukawa model -- if reduced to the UV-complete trajectory -- thus corresponds to
initial conditions of the fermionic model in the perturbative weak coupling
regime.

\section{Conclusions}
\label{sec:conc}

The relativistic Yukawa model proposed in this work exhibits features that are
both novel and, to the best of our knowledge, unprecedented in quantum field
theories in four-dimensional spacetime. In contrast to conventional models
involving Dirac, Majorana or Weyl fermions, the use of relativistic Luttinger
fermions exerts a strong influence on the RG phase diagram of the model in the
space spanned by the power-counting RG relevant and marginal couplings: for
generic initial conditions (including those with an arbitrarily positive scalar
mass parameter in units of the cutoff scale), the model features an IR
attractive partial fixed point of the RG evolution at which the system can
bridge a wide range of scales. While all couplings are RG irrelevant at the
fixed point, the scalar mass parameter is RG marginal and exhibits a slow
logarithmic drift towards small values. 

The long-range behavior of the model is characterized by spontaneous symmetry
breaking and mass gap generation in both the scalar and the fermionic sector.
Remarkably, the low-energy scales such as the scalar condensate or mass gaps can
be many orders of magnitude smaller than a microscopic UV cutoff scale without
the need to fine-tune initial parameters. In fact, the UV and IR scales are
naturally many orders of magnitude apart for generic initial conditions,
including those with couplings of order one and a scalar mass parameter of the
order of several times the UV cutoff scale. 

Some of these exceptional properties of our Yukawa model are reminiscent to the
phenomenon of self-organized criticality in statistical or dynamical systems:
identifying the RG time with the physical time in dynamical systems, our model
inevitably runs towards a scale where it becomes critical in the sense of an
onset of spontaneous symmetry breaking. At this scale, all modes are gapless
featuring large fluctuations. For generic initial conditions, the long-range
behavior is universal because of the IR attractiveness of the partial fixed
point at which the dependence of the system on its initial condition is depleted
and largely removed. The partial fixed point is characterized by critical
exponents governing the RG running of the dimensionless couplings in terms of
simple power-laws. The most prominent similarity to self-organized criticality
is given by the slow logarithmic running of the (dimensionless) scalar mass
parameter which plays the role of a slow driving force that gradually and
inevitably moves the model to criticality.  

The present model therefore is a concrete realization of a scenario
envisaged in \cite{Bornholdt:1992up} for addressing the issue of naturalness in
elementary particle physics in terms of self-organized criticality. Whether or
not the present mechanism can be used for corresponding model building in
elementary particle physics is an open question. Possible pathways include
adding a separate Luttinger fermionic sector to the standard model, or embedding
its fermionic content into Luttinger spinors; for a first assessment see
\cite{Gies:2023cnd}. While speculative, it might be an inspiring observation to
see that the universal scalar-to-fermion mass gap ratio of our model is
quantitatively close to the Higgs-to-top mass ratio in the standard model. 

Within the functional RG approach, we have been able to derive this mass-gap
ratio together with a number of quantitative results for the mass spectrum.
Provided the model spends sufficient RG time near the partial fixed point, as is
the case for generic initial conditions, the long-range properties of the model
are universal. While some of our nonperturbative results can also be verified
with large-$\Nf$ techniques as used in this work as well, we believe that these
quantitative long-range properties could be a prime example for the application
of other techniques such Dyson-Schwinger or gap equations, or lattice field
theory. Such techniques can also help shedding further light on the nature of the fermionic mass gap in our model which occurs in the form of a pair of  (Minkowskian) complex conjugate poles in the complex $p^2$ plane and the question of the existence or inexistence of Luttinger fermions as asymptotic states \cite{Gies:2024dly}. 

Finally, an attractive feature of our model is that it features a UV complete
extension by virtue of the asymptotically free purely fermionic model which is
in the same universality class as our Yukawa model. Though our results for the Yukawa model are independent of this possible UV completion, the existence of such a scale-invariant high-energy limit may represent another motivation to explore such models with relativistic Luttinger fermions even further.

%%%%%%%%%%%%%%%%%%%%%%%%%%%%%%%%%%%%%%%%
\section*{Acknowledgments}
%%%%%%%%%%%%%%%%%%%%%%%%%%%%%%%%%%%%%%%%

We thank Lukas Janssen, David Moser, Richard Schmieden, and Christof Wetterich
for valuable discussions. We are grateful to Richard Schmieden for his kind
assistance with the numerical implementation and the code developement. HG
thanks the ITP Heidelberg for hospitality while working on this project. This
work has been funded by the Deutsche Forschungsgemeinschaft (DFG) under Grant
No. 406116891 within the Research Training Group RTG 2522/1.

\appendix
\setcounter{equation}{0}
\setcounter{figure}{0}
\setcounter{table}{0}
\makeatletter
\renewcommand{\theequation}{A\arabic{equation}}
\renewcommand{\thefigure}{A\arabic{figure}}
\setcounter{secnumdepth}{3}

\section{Abrikosov algebra}
\label{sec:AppA}

For completeness, we recall a few aspects of the Abrikosov algebra
\cite{Abrikosov:1974a} in \Eqref{eq:AbrikosovA} in a relativistic context as
studied in \cite{Gies:2023cnd,Gies:2024dly}. 

While not explicitly needed, it is helpful to know that a representation of the
$G_{\mu\nu}$ matrices can be constructed in terms of a Euclidean Dirac algebra
$\{\gamma_A,\gamma_B\} = 2 \delta_{AB}$. For the present work, we work in four-dimensional Euclidean spacetime with metric $g=\text{diag}(1,1,1,1)$, such that the Abrikosov algebra is satisfied by
\begin{eqnarray}
 G_{0i}&=&- \sqrt{\frac{2}{3}} \gamma_{A=i}, \quad i=1,2,3, \nonumber\\
 G_{12}&=& -\sqrt{\frac{2}{3}} \gamma_4, \quad G_{23} = -\sqrt{\frac{2}{3}} 
\gamma_5, \quad G_{31}= -\sqrt{\frac{2}{3}} \gamma_6, \nonumber\\
G_{00}&=& - \gamma_7, \quad G_{11} = - \frac{1}{3} \gamma_7 
- \frac{2\sqrt{2}}{3} 
\gamma_8,\label{eq:Grep}\\
G_{22}&=& -\frac{1}{3} \gamma_7 + \frac{\sqrt{2}}{3} \gamma_8 - 
\sqrt{\frac{2}{3}} \gamma_9, \nonumber\\
G_{33}&=& -\frac{1}{3} \gamma_7 + \frac{\sqrt{2}}{3} \gamma_8 + 
\sqrt{\frac{2}{3}} \gamma_9, \nonumber
\end{eqnarray}
in agreement with the Euclidean rotation of the Minkowskian version discussed in
\cite{Gies:2024dly}. This representation can be related to that of
\cite{Janssen:2015xga} for $d=4$ Euclidean dimensions through a spin-base
transformation. While \Eqref{eq:Grep} can be constructed from 9 Euclidean Dirac
matrices $\gamma_{1,\dots,9}$, a Euclidean construction satisfying also
reflection positivity requires a nontrivial spin metric $h$ for the definition
of a Luttinger conjugate spinor $\bar\psi= \psi^\dagger h$. As detailed in
\cite{Gies:2023cnd,Gies:2024dly}, this demands for an at least $d_\gamma=32$
dimensional representation of the Euclidean Dirac algebra, going along with two
further anti-commuting elements $\gamma_{10}$ and $\gamma_{11}$. In the present
work, we use $\gamma_{10}$ for the construction of the spin metric, $h=\gamma_1
\gamma_2 \gamma_3 \gamma_{10}$, while $\gamma_{11}$ is employed for the
definition of the Yukawa interaction of our model. Various other choices would
alternatively be possible, cf. \cite{Gies:2024dly}.

\section{Threshold functions}
\label{app:thresh}

\setcounter{equation}{1}

The threshold functions used in the main text, are classified and defined for general regulator functions in the literature \cite{Jungnickel:1995fp,Berges:2000ew,Hofling:2002hj,Gies:2013pma,Gies:2020xuh}. As the Luttinger fermions come with a new kinetic term, several new threshold functions are needed which can be defined in full analogy to those involving, for instance, Dirac fermions.

For the regulator, we choose in the scalar and the fermionic sectors
\begin{eqnarray}
    R_{k,\phi}(p^2) &=& Z_\phi p^2 (1+ r_\text{B}(p^2/k^2)), \nonumber\\ 
    R_{k,\psi}(p^2) &=& Z_\psi G_{\mu\nu} p^\mu p^\nu (1+ r_\text{L}(p^2/k^2)),
 \label{eq:regfun}
\end{eqnarray}
where $r_\text{B}(y),r_{\text{L}}(y)$ denote dimensionless regulator shape functions that encode the momentum space regularization near $p^2\sim k^2$. Introducing the following auxiliary functions related to  the regularized momentum-space propagators
\begin{equation}
    G_{\text{B}}(\omega)= \frac{1}{y(1+r_{\text{B}})+\omega}, \,
    G_{\text{L}}(\omega)= \frac{1}{y^2(1+r_{\text{L}})^2+\omega},
\end{equation}
where $y=p^2/k^2$, the threshold functions involving Luttinger fermions occuring
in the main text are defined by 
\begin{eqnarray}
&&    l_{0}^{(\text{L})\,d}(\omega ; \eta) =  \frac{k^{-d}}{4 v_d} \int_p \tilde\partial_t \ln G_{\text{L}}^{-1}(\omega) , \label{eq:lL0}\\
&&    l_{1,1}^{(\text{LB})\,d}(\omega_{1},\omega_{2};
         \eta_{1},\eta_{2})=  \frac{k^{-d}}{4 v_d} \int_p \tilde\partial_t G_{\text{L}}(\omega_1) G_{\text{B}}(\omega_2),\label{eq:lL11}\\
&& m_{1,2}^{(\text{LB}),d}(\omega_1,\omega_2;\eta_{1},\eta_{2}) 
        =-\frac{k^{-d}}{4v_d} \int_p dy \ y^2 \nonumber\\
        &&\qquad\qquad\tilde\partial_t [(G_{\text{B}} (\omega_2))''(1+r_\text{L}) G_{\text{L}} (\omega_1)]  ,\label{eq:mLB12} \\
&&	  m_4^{(\text{L}),d}(\omega,\eta)= -\frac{k^{-d}}{4v_d} \int_p\ y \ \tilde\partial_t \biggl\{ \left[\left( y(1+r_L) G_{\text{L}} (\omega)\right)'\right]^2 \nonumber\\
  &&\qquad\qquad +\frac{d}{2}(1+r_{\text{L}})^2 G^2_{\text{L}}(\omega) \biggr\}  , \label{eq:mL4}\\
&&   m_2^{(\text{L}),d}(\omega,\eta)= -\frac{k^{-d}}{4v_d} \int_p \ y \ \tilde\partial_t [(G_{\text{L}} (\omega))']^2 ,   \label{eq:mL2}
\end{eqnarray}
where, in practice, the derivative $\tilde\partial_t$ can be read as
$\tilde\partial_t\to(\partial_t r -\eta r)\partial_r$, $\int_p\equiv
\int\frac{d^dp}{(2\pi)^d}$ indicates the full momentum integral, and primes
denote derivatives with respect to $y$.

For all concrete computations in the main text, we use the partially linear
regulator (Litim regulator) \cite{Litim:2000ci,Litim:2001up},
\begin{equation}
    r_{\text{B}}=r_{\text{L}}= \left(\frac{1}{y}-1\right) \theta(1-y),
    \label{eq:Litim}
\end{equation}
which allows for an analytic evaluation of the loop momentum integration. The corresponding threshold functions then read
\begin{align}
& l_{0}^{(\text{L})\,d}(\omega ; \eta) = \frac{4}{d} \left(1-\frac{\eta}{d+2} \right) \frac{1}{1+\omega} , \label{eq:lL0l}\\
& l_{1,1}^{(\text{LB})\,d}(\omega_{1},\omega_{2};
         \eta_{1},\eta_{2})= \frac{2}{d} \frac{1}{(1+\omega_1)(1+\omega_2)} \nonumber\\
         &\qquad \qquad \times \left(\frac{2}{1+\omega_1} \left(1-\frac{\eta_1}{d+2} \right) \right. \nonumber\\
         &\qquad \qquad + \left.\frac{1}{1+\omega_2} \left(1-\frac{\eta_2}{d+2} \right) \right)  ,       \label{eq:lL11l}\\
& m_{1,2}^{(\text{LB}),d}(\omega_1,\omega_2;\eta_{1},\eta_{2}) 
        = \frac{1}{2} \frac{1}{(1+\omega_1)(1+\omega_2)^2} \nonumber\\
        &\qquad \qquad \times \left( d+1 + \frac{\omega_1-3}{\omega_1 +1} -\eta_2 \right) ,\label{eq:mLB12l}\\ 
& m_4^{(\text{L}),d}(\omega,\eta)= \frac{(1-\omega)^2}{(1+\omega)^4} + \frac{2(d-\eta)}{d-2}  \frac{1-\omega}{(1+\omega)^3} ,  \label{eq:mL4l}\\
&m_2^{(\text{L}),d}(\omega,\eta)= \frac{4}{(1+\omega)^4} .   \label{eq:mL2l}
\end{align}
For completeness, we also list all other required threshold functions known from the literature \cite{Jungnickel:1995fp,Berges:2000ew,Hofling:2002hj,Gies:2013pma,Gies:2020xuh}:
\begin{align}
& l_{0}^{d}(\omega ; \eta) = \frac{2}{d} \left(1- \frac{\eta}{d+2}\right) \frac{1}{1+\omega}, \label{eq:l0}\\
& m_{2,2}^{d}(\omega;\eta) = \frac{1}{(1+\omega)^4}. \label{eq:m22}
\end{align}
Because of the nonanalyticity of the Litim regulator, the $m$-type threshold functions partly involve ill-defined products of distributions such as $\delta(1-y) \theta(1-y)$. This is a result of using the derivative expansion of the action as an ansatz; including full momentum-dependencies would lead to perfectly well-defined flows. In the present case, the problematic products occuring here can straightforwardly be cured by suitably smearing the singularity of the Heaviside function. Using a symmetric smearing, it can be shown that the resulting loop integrations can effectively be performed by the simple replacement $\delta(1-y) \theta(1-y) \to \frac{1}{2} \delta (1-y)$. This recipe is in agreement with the results used in the literature.

\bibliography{bibliography}

\begin{thebibliography}{66}%
\makeatletter
\providecommand \@ifxundefined [1]{%
 \@ifx{#1\undefined}
}%
\providecommand \@ifnum [1]{%
 \ifnum #1\expandafter \@firstoftwo
 \else \expandafter \@secondoftwo
 \fi
}%
\providecommand \@ifx [1]{%
 \ifx #1\expandafter \@firstoftwo
 \else \expandafter \@secondoftwo
 \fi
}%
\providecommand \natexlab [1]{#1}%
\providecommand \enquote  [1]{``#1''}%
\providecommand \bibnamefont  [1]{#1}%
\providecommand \bibfnamefont [1]{#1}%
\providecommand \citenamefont [1]{#1}%
\providecommand \href@noop [0]{\@secondoftwo}%
\providecommand \href [0]{\begingroup \@sanitize@url \@href}%
\providecommand \@href[1]{\@@startlink{#1}\@@href}%
\providecommand \@@href[1]{\endgroup#1\@@endlink}%
\providecommand \@sanitize@url [0]{\catcode `\\12\catcode `\$12\catcode
  `\&12\catcode `\#12\catcode `\^12\catcode `\_12\catcode `\%12\relax}%
\providecommand \@@startlink[1]{}%
\providecommand \@@endlink[0]{}%
\providecommand \url  [0]{\begingroup\@sanitize@url \@url }%
\providecommand \@url [1]{\endgroup\@href {#1}{\urlprefix }}%
\providecommand \urlprefix  [0]{URL }%
\providecommand \Eprint [0]{\href }%
\providecommand \doibase [0]{http://dx.doi.org/}%
\providecommand \selectlanguage [0]{\@gobble}%
\providecommand \bibinfo  [0]{\@secondoftwo}%
\providecommand \bibfield  [0]{\@secondoftwo}%
\providecommand \translation [1]{[#1]}%
\providecommand \BibitemOpen [0]{}%
\providecommand \bibitemStop [0]{}%
\providecommand \bibitemNoStop [0]{.\EOS\space}%
\providecommand \EOS [0]{\spacefactor3000\relax}%
\providecommand \BibitemShut  [1]{\csname bibitem#1\endcsname}%
\let\auto@bib@innerbib\@empty
%</preamble>
\bibitem [{\citenamefont {Zinn-Justin}(1989)}]{ZinnJustin:1989mi}%
  \BibitemOpen
  \bibfield  {author} {\bibinfo {author} {\bibfnamefont {J.}~\bibnamefont
  {Zinn-Justin}},\ }\href@noop {} {\bibfield  {journal} {\bibinfo  {journal}
  {Int. Ser. Monogr. Phys.}\ }\textbf {\bibinfo {volume} {77}},\ \bibinfo
  {pages} {1} (\bibinfo {year} {1989})}\BibitemShut {NoStop}%
%%CITATION = IMPHA,77,1;%%
\bibitem [{\citenamefont {'t~Hooft}(1980)}]{tHooft:1979rat}%
  \BibitemOpen
  \bibfield  {author} {\bibinfo {author} {\bibfnamefont {G.}~\bibnamefont
  {'t~Hooft}},\ }\href {\doibase 10.1007/978-1-4684-7571-5_9} {\bibfield
  {journal} {\bibinfo  {journal} {NATO Sci. Ser. B}\ }\textbf {\bibinfo
  {volume} {59}},\ \bibinfo {pages} {135} (\bibinfo {year} {1980})}\BibitemShut
  {NoStop}%
\bibitem [{\citenamefont {Giudice}(2008)}]{Giudice:2008bi}%
  \BibitemOpen
  \bibfield  {author} {\bibinfo {author} {\bibfnamefont {G.~F.}\ \bibnamefont
  {Giudice}},\ }\href {\doibase 10.1142/9789812779762_0010} {\ ,\ \bibinfo
  {pages} {155} (\bibinfo {year} {2008})},\ \Eprint
  {http://arxiv.org/abs/0801.2562} {arXiv:0801.2562 [hep-ph]} \BibitemShut
  {NoStop}%
\bibitem [{\citenamefont {Grinstein}\ \emph {et~al.}(2008)\citenamefont
  {Grinstein}, \citenamefont {O'Connell},\ and\ \citenamefont
  {Wise}}]{Grinstein:2007mp}%
  \BibitemOpen
  \bibfield  {author} {\bibinfo {author} {\bibfnamefont {B.}~\bibnamefont
  {Grinstein}}, \bibinfo {author} {\bibfnamefont {D.}~\bibnamefont
  {O'Connell}}, \ and\ \bibinfo {author} {\bibfnamefont {M.~B.}\ \bibnamefont
  {Wise}},\ }\href {\doibase 10.1103/PhysRevD.77.025012} {\bibfield  {journal}
  {\bibinfo  {journal} {Phys. Rev. D}\ }\textbf {\bibinfo {volume} {77}},\
  \bibinfo {pages} {025012} (\bibinfo {year} {2008})},\ \Eprint
  {http://arxiv.org/abs/0704.1845} {arXiv:0704.1845 [hep-ph]} \BibitemShut
  {NoStop}%
\bibitem [{\citenamefont {Hossenfelder}(2021)}]{Hossenfelder:2018ikr}%
  \BibitemOpen
  \bibfield  {author} {\bibinfo {author} {\bibfnamefont {S.}~\bibnamefont
  {Hossenfelder}},\ }\href {\doibase 10.1007/s11229-019-02377-5} {\bibfield
  {journal} {\bibinfo  {journal} {Synthese}\ }\textbf {\bibinfo {volume}
  {198}},\ \bibinfo {pages} {3727} (\bibinfo {year} {2021})},\ \Eprint
  {http://arxiv.org/abs/1801.02176} {arXiv:1801.02176 [physics.hist-ph]}
  \BibitemShut {NoStop}%
\bibitem [{\citenamefont {Bak}\ \emph {et~al.}(1987)\citenamefont {Bak},
  \citenamefont {Tang},\ and\ \citenamefont {Wiesenfeld}}]{Bak:1987xua}%
  \BibitemOpen
  \bibfield  {author} {\bibinfo {author} {\bibfnamefont {P.}~\bibnamefont
  {Bak}}, \bibinfo {author} {\bibfnamefont {C.}~\bibnamefont {Tang}}, \ and\
  \bibinfo {author} {\bibfnamefont {K.}~\bibnamefont {Wiesenfeld}},\ }\href
  {\doibase 10.1103/PhysRevLett.59.381} {\bibfield  {journal} {\bibinfo
  {journal} {Phys. Rev. Lett.}\ }\textbf {\bibinfo {volume} {59}},\ \bibinfo
  {pages} {381} (\bibinfo {year} {1987})}\BibitemShut {NoStop}%
\bibitem [{\citenamefont {Bak}\ \emph {et~al.}(1988)\citenamefont {Bak},
  \citenamefont {Tang},\ and\ \citenamefont {Wiesenfeld}}]{Bak:1988zz}%
  \BibitemOpen
  \bibfield  {author} {\bibinfo {author} {\bibfnamefont {P.}~\bibnamefont
  {Bak}}, \bibinfo {author} {\bibfnamefont {C.}~\bibnamefont {Tang}}, \ and\
  \bibinfo {author} {\bibfnamefont {K.}~\bibnamefont {Wiesenfeld}},\ }\href
  {\doibase 10.1103/PhysRevA.38.364} {\bibfield  {journal} {\bibinfo  {journal}
  {Phys. Rev. A}\ }\textbf {\bibinfo {volume} {38}},\ \bibinfo {pages} {364}
  (\bibinfo {year} {1988})}\BibitemShut {NoStop}%
\bibitem [{\citenamefont {Manna}(1991)}]{Manna:1991}%
  \BibitemOpen
  \bibfield  {author} {\bibinfo {author} {\bibfnamefont {S.~S.}\ \bibnamefont
  {Manna}},\ }\href {\doibase 10.1088/0305-4470/24/7/009} {\bibfield  {journal}
  {\bibinfo  {journal} {Journal of Physics A: Mathematical and General}\
  }\textbf {\bibinfo {volume} {24}},\ \bibinfo {pages} {L363} (\bibinfo {year}
  {1991})}\BibitemShut {NoStop}%
\bibitem [{\citenamefont {Olami}\ \emph {et~al.}(1992)\citenamefont {Olami},
  \citenamefont {Feder},\ and\ \citenamefont {Christensen}}]{Olami:1992}%
  \BibitemOpen
  \bibfield  {author} {\bibinfo {author} {\bibfnamefont {Z.}~\bibnamefont
  {Olami}}, \bibinfo {author} {\bibfnamefont {H.~J.~S.}\ \bibnamefont {Feder}},
  \ and\ \bibinfo {author} {\bibfnamefont {K.}~\bibnamefont {Christensen}},\
  }\href {\doibase 10.1103/PhysRevLett.68.1244} {\bibfield  {journal} {\bibinfo
   {journal} {Phys. Rev. Lett.}\ }\textbf {\bibinfo {volume} {68}},\ \bibinfo
  {pages} {1244} (\bibinfo {year} {1992})}\BibitemShut {NoStop}%
\bibitem [{\citenamefont {Malamud}\ \emph {et~al.}(1998)\citenamefont
  {Malamud}, \citenamefont {Morein},\ and\ \citenamefont
  {Turcotte}}]{Malamud:1998}%
  \BibitemOpen
  \bibfield  {author} {\bibinfo {author} {\bibfnamefont {B.~D.}\ \bibnamefont
  {Malamud}}, \bibinfo {author} {\bibfnamefont {G.}~\bibnamefont {Morein}}, \
  and\ \bibinfo {author} {\bibfnamefont {D.~L.}\ \bibnamefont {Turcotte}},\
  }\href {\doibase 10.1126/science.281.5384.1840} {\bibfield  {journal}
  {\bibinfo  {journal} {Science}\ }\textbf {\bibinfo {volume} {281}},\ \bibinfo
  {pages} {1840} (\bibinfo {year} {1998})}\BibitemShut {NoStop}%
\bibitem [{\citenamefont {Vespignani}\ and\ \citenamefont
  {Zapperi}(1998)}]{Vespignani:1998}%
  \BibitemOpen
  \bibfield  {author} {\bibinfo {author} {\bibfnamefont {A.}~\bibnamefont
  {Vespignani}}\ and\ \bibinfo {author} {\bibfnamefont {S.}~\bibnamefont
  {Zapperi}},\ }\href {\doibase 10.1103/physreve.57.6345} {\bibfield  {journal}
  {\bibinfo  {journal} {Physical Review E}\ }\textbf {\bibinfo {volume} {57}},\
  \bibinfo {pages} {6345–6362} (\bibinfo {year} {1998})}\BibitemShut
  {NoStop}%
\bibitem [{\citenamefont {Bornholdt}\ and\ \citenamefont
  {Wetterich}(1992)}]{Bornholdt:1992up}%
  \BibitemOpen
  \bibfield  {author} {\bibinfo {author} {\bibfnamefont {S.}~\bibnamefont
  {Bornholdt}}\ and\ \bibinfo {author} {\bibfnamefont {C.}~\bibnamefont
  {Wetterich}},\ }\href {\doibase 10.1016/0370-2693(92)90659-R} {\bibfield
  {journal} {\bibinfo  {journal} {Phys. Lett. B}\ }\textbf {\bibinfo {volume}
  {282}},\ \bibinfo {pages} {399} (\bibinfo {year} {1992})}\BibitemShut
  {NoStop}%
\bibitem [{\citenamefont {Gies}\ \emph {et~al.}(2024)\citenamefont {Gies},
  \citenamefont {Heinzel}, \citenamefont {Laufk\"otter},\ and\ \citenamefont
  {Picciau}}]{Gies:2023cnd}%
  \BibitemOpen
  \bibfield  {author} {\bibinfo {author} {\bibfnamefont {H.}~\bibnamefont
  {Gies}}, \bibinfo {author} {\bibfnamefont {P.}~\bibnamefont {Heinzel}},
  \bibinfo {author} {\bibfnamefont {J.}~\bibnamefont {Laufk\"otter}}, \ and\
  \bibinfo {author} {\bibfnamefont {M.}~\bibnamefont {Picciau}},\ }\href
  {\doibase 10.1103/PhysRevD.110.065001} {\bibfield  {journal} {\bibinfo
  {journal} {Phys. Rev. D}\ }\textbf {\bibinfo {volume} {110}},\ \bibinfo
  {pages} {065001} (\bibinfo {year} {2024})},\ \Eprint
  {http://arxiv.org/abs/2312.12058} {arXiv:2312.12058 [hep-th]} \BibitemShut
  {NoStop}%
\bibitem [{\citenamefont {Luttinger}(1956)}]{LuttingerPhysRev.102.1030}%
  \BibitemOpen
  \bibfield  {author} {\bibinfo {author} {\bibfnamefont {J.~M.}\ \bibnamefont
  {Luttinger}},\ }\href {\doibase 10.1103/PhysRev.102.1030} {\bibfield
  {journal} {\bibinfo  {journal} {Phys. Rev.}\ }\textbf {\bibinfo {volume}
  {102}},\ \bibinfo {pages} {1030} (\bibinfo {year} {1956})}\BibitemShut
  {NoStop}%
\bibitem [{\citenamefont {Murakami}\ \emph {et~al.}(2004)\citenamefont
  {Murakami}, \citenamefont {Nagosa},\ and\ \citenamefont
  {Zhang}}]{Murakami:2004zz}%
  \BibitemOpen
  \bibfield  {author} {\bibinfo {author} {\bibfnamefont {S.}~\bibnamefont
  {Murakami}}, \bibinfo {author} {\bibfnamefont {N.}~\bibnamefont {Nagosa}}, \
  and\ \bibinfo {author} {\bibfnamefont {S.-C.}\ \bibnamefont {Zhang}},\ }\href
  {\doibase 10.1103/PhysRevB.69.235206} {\bibfield  {journal} {\bibinfo
  {journal} {Phys. Rev. B}\ }\textbf {\bibinfo {volume} {69}},\ \bibinfo
  {pages} {235206} (\bibinfo {year} {2004})},\ \Eprint
  {http://arxiv.org/abs/cond-mat/0310005} {arXiv:cond-mat/0310005} \BibitemShut
  {NoStop}%
\bibitem [{\citenamefont {Moon}\ \emph {et~al.}(2013)\citenamefont {Moon},
  \citenamefont {Xu}, \citenamefont {Kim},\ and\ \citenamefont
  {Balents}}]{Moon:2012rx}%
  \BibitemOpen
  \bibfield  {author} {\bibinfo {author} {\bibfnamefont {E.-G.}\ \bibnamefont
  {Moon}}, \bibinfo {author} {\bibfnamefont {C.}~\bibnamefont {Xu}}, \bibinfo
  {author} {\bibfnamefont {Y.~B.}\ \bibnamefont {Kim}}, \ and\ \bibinfo
  {author} {\bibfnamefont {L.}~\bibnamefont {Balents}},\ }\href {\doibase
  10.1103/PhysRevLett.111.206401} {\bibfield  {journal} {\bibinfo  {journal}
  {Phys. Rev. Lett.}\ }\textbf {\bibinfo {volume} {111}},\ \bibinfo {pages}
  {206401} (\bibinfo {year} {2013})},\ \Eprint {http://arxiv.org/abs/1212.1168}
  {arXiv:1212.1168 [cond-mat.str-el]} \BibitemShut {NoStop}%
\bibitem [{\citenamefont {Janssen}\ and\ \citenamefont
  {Herbut}(2015)}]{Janssen:2015xga}%
  \BibitemOpen
  \bibfield  {author} {\bibinfo {author} {\bibfnamefont {L.}~\bibnamefont
  {Janssen}}\ and\ \bibinfo {author} {\bibfnamefont {I.~F.}\ \bibnamefont
  {Herbut}},\ }\href {\doibase 10.1103/PhysRevB.92.045117} {\bibfield
  {journal} {\bibinfo  {journal} {Phys. Rev. B}\ }\textbf {\bibinfo {volume}
  {92}},\ \bibinfo {pages} {045117} (\bibinfo {year} {2015})},\ \Eprint
  {http://arxiv.org/abs/1503.04242} {arXiv:1503.04242 [cond-mat.str-el]}
  \BibitemShut {NoStop}%
\bibitem [{\citenamefont {Janssen}\ and\ \citenamefont
  {Herbut}(2017)}]{Janssen:2016xvc}%
  \BibitemOpen
  \bibfield  {author} {\bibinfo {author} {\bibfnamefont {L.}~\bibnamefont
  {Janssen}}\ and\ \bibinfo {author} {\bibfnamefont {I.~F.}\ \bibnamefont
  {Herbut}},\ }\href {\doibase 10.1103/PhysRevB.95.075101} {\bibfield
  {journal} {\bibinfo  {journal} {Phys. Rev. B}\ }\textbf {\bibinfo {volume}
  {95}},\ \bibinfo {pages} {075101} (\bibinfo {year} {2017})},\ \Eprint
  {http://arxiv.org/abs/1611.04594} {arXiv:1611.04594 [cond-mat.str-el]}
  \BibitemShut {NoStop}%
\bibitem [{\citenamefont {Ray}\ \emph {et~al.}(2018)\citenamefont {Ray},
  \citenamefont {Vojta},\ and\ \citenamefont {Janssen}}]{Ray:2018gtp}%
  \BibitemOpen
  \bibfield  {author} {\bibinfo {author} {\bibfnamefont {S.}~\bibnamefont
  {Ray}}, \bibinfo {author} {\bibfnamefont {M.}~\bibnamefont {Vojta}}, \ and\
  \bibinfo {author} {\bibfnamefont {L.}~\bibnamefont {Janssen}},\ }\href
  {\doibase 10.1103/PhysRevB.98.245128} {\bibfield  {journal} {\bibinfo
  {journal} {Phys. Rev. B}\ }\textbf {\bibinfo {volume} {98}},\ \bibinfo
  {pages} {245128} (\bibinfo {year} {2018})},\ \Eprint
  {http://arxiv.org/abs/1810.07695} {arXiv:1810.07695 [cond-mat.str-el]}
  \BibitemShut {NoStop}%
\bibitem [{\citenamefont {Dey}\ and\ \citenamefont
  {Maciejko}(2022)}]{Dey:2022lkx}%
  \BibitemOpen
  \bibfield  {author} {\bibinfo {author} {\bibfnamefont {S.}~\bibnamefont
  {Dey}}\ and\ \bibinfo {author} {\bibfnamefont {J.}~\bibnamefont {Maciejko}},\
  }\href {\doibase 10.1103/PhysRevB.106.035140} {\bibfield  {journal} {\bibinfo
   {journal} {Phys. Rev. B}\ }\textbf {\bibinfo {volume} {106}},\ \bibinfo
  {pages} {035140} (\bibinfo {year} {2022})},\ \Eprint
  {http://arxiv.org/abs/2204.05319} {arXiv:2204.05319 [cond-mat.str-el]}
  \BibitemShut {NoStop}%
\bibitem [{\citenamefont {Moser}\ and\ \citenamefont
  {Janssen}(2024)}]{Moser:2024dmq}%
  \BibitemOpen
  \bibfield  {author} {\bibinfo {author} {\bibfnamefont {D.~J.}\ \bibnamefont
  {Moser}}\ and\ \bibinfo {author} {\bibfnamefont {L.}~\bibnamefont
  {Janssen}},\ }\href@noop {} {\  (\bibinfo {year} {2024})},\ \Eprint
  {http://arxiv.org/abs/2412.06890} {arXiv:2412.06890 [cond-mat.str-el]}
  \BibitemShut {NoStop}%
\bibitem [{\citenamefont {Gies}\ and\ \citenamefont
  {Picciau}(2025)}]{Gies:2024dly}%
  \BibitemOpen
  \bibfield  {author} {\bibinfo {author} {\bibfnamefont {H.}~\bibnamefont
  {Gies}}\ and\ \bibinfo {author} {\bibfnamefont {M.}~\bibnamefont {Picciau}},\
  }\href {\doibase 10.1103/PhysRevD.111.085001} {\bibfield  {journal} {\bibinfo
   {journal} {Phys. Rev. D}\ }\textbf {\bibinfo {volume} {111}},\ \bibinfo
  {pages} {085001} (\bibinfo {year} {2025})},\ \Eprint
  {http://arxiv.org/abs/2410.22166} {arXiv:2410.22166 [hep-th]} \BibitemShut
  {NoStop}%
\bibitem [{\citenamefont {Abrikosov}(1974)}]{Abrikosov:1974a}%
  \BibitemOpen
  \bibfield  {author} {\bibinfo {author} {\bibfnamefont {A.~A.}\ \bibnamefont
  {Abrikosov}},\ }\href
  {http://www.jetp.ras.ru/cgi-bin/e/index/e/39/4/p709?a=list} {\bibfield
  {journal} {\bibinfo  {journal} {Sov. Phys. JETP}\ }\textbf {\bibinfo {volume}
  {39}},\ \bibinfo {pages} {709} (\bibinfo {year} {1974})}\BibitemShut
  {NoStop}%
\bibitem [{\citenamefont {Holland}\ and\ \citenamefont
  {Kuti}(2004)}]{Holland:2003jr}%
  \BibitemOpen
  \bibfield  {author} {\bibinfo {author} {\bibfnamefont {K.}~\bibnamefont
  {Holland}}\ and\ \bibinfo {author} {\bibfnamefont {J.}~\bibnamefont {Kuti}},\
  }\bibfield  {booktitle} {\emph {\bibinfo {booktitle} {{Lattice hadron
  physics. Proceedings, 2nd Topical Workshop, LHP 2003, Cairns, Australia, July
  22-30, 2003}}},\ }\href {\doibase 10.1016/S0920-5632(03)02706-3} {\bibfield
  {journal} {\bibinfo  {journal} {Nucl. Phys. Proc. Suppl.}\ }\textbf {\bibinfo
  {volume} {129}},\ \bibinfo {pages} {765} (\bibinfo {year} {2004})},\ \bibinfo
  {note} {[,765(2003)]},\ \Eprint {http://arxiv.org/abs/hep-lat/0308020}
  {arXiv:hep-lat/0308020 [hep-lat]} \BibitemShut {NoStop}%
%%CITATION = HEP-LAT/0308020;%%
\bibitem [{\citenamefont {Holland}(2005)}]{Holland:2004sd}%
  \BibitemOpen
  \bibfield  {author} {\bibinfo {author} {\bibfnamefont {K.}~\bibnamefont
  {Holland}},\ }\bibfield  {booktitle} {\emph {\bibinfo {booktitle} {{Lattice
  field theory. Proceedings, 22nd International Symposium, Lattice 2004,
  Batavia, USA, June 21-26, 2004}}},\ }\href {\doibase
  10.1016/j.nuclphysbps.2004.11.293} {\bibfield  {journal} {\bibinfo  {journal}
  {Nucl. Phys. Proc. Suppl.}\ }\textbf {\bibinfo {volume} {140}},\ \bibinfo
  {pages} {155} (\bibinfo {year} {2005})},\ \bibinfo {note} {[,155(2004)]},\
  \Eprint {http://arxiv.org/abs/hep-lat/0409112} {arXiv:hep-lat/0409112
  [hep-lat]} \BibitemShut {NoStop}%
%%CITATION = HEP-LAT/0409112;%%
\bibitem [{\citenamefont {Branchina}\ and\ \citenamefont
  {Faivre}(2005)}]{Branchina:2005tu}%
  \BibitemOpen
  \bibfield  {author} {\bibinfo {author} {\bibfnamefont {V.}~\bibnamefont
  {Branchina}}\ and\ \bibinfo {author} {\bibfnamefont {H.}~\bibnamefont
  {Faivre}},\ }\href {\doibase 10.1103/PhysRevD.72.065017} {\bibfield
  {journal} {\bibinfo  {journal} {Phys. Rev.}\ }\textbf {\bibinfo {volume}
  {D72}},\ \bibinfo {pages} {065017} (\bibinfo {year} {2005})},\ \Eprint
  {http://arxiv.org/abs/hep-th/0503188} {arXiv:hep-th/0503188 [hep-th]}
  \BibitemShut {NoStop}%
%%CITATION = HEP-TH/0503188;%%
\bibitem [{\citenamefont {Branchina}\ \emph {et~al.}(2009)\citenamefont
  {Branchina}, \citenamefont {Faivre},\ and\ \citenamefont
  {Pangon}}]{Branchina:2008pc}%
  \BibitemOpen
  \bibfield  {author} {\bibinfo {author} {\bibfnamefont {V.}~\bibnamefont
  {Branchina}}, \bibinfo {author} {\bibfnamefont {H.}~\bibnamefont {Faivre}}, \
  and\ \bibinfo {author} {\bibfnamefont {V.}~\bibnamefont {Pangon}},\ }\href
  {\doibase 10.1088/0954-3899/36/1/015006} {\bibfield  {journal} {\bibinfo
  {journal} {J. Phys.}\ }\textbf {\bibinfo {volume} {G36}},\ \bibinfo {pages}
  {015006} (\bibinfo {year} {2009})},\ \Eprint {http://arxiv.org/abs/0802.4423}
  {arXiv:0802.4423 [hep-ph]} \BibitemShut {NoStop}%
%%CITATION = ARXIV:0802.4423;%%
\bibitem [{\citenamefont {Gies}\ \emph {et~al.}(2014)\citenamefont {Gies},
  \citenamefont {Gneiting},\ and\ \citenamefont {Sondenheimer}}]{Gies:2013fua}%
  \BibitemOpen
  \bibfield  {author} {\bibinfo {author} {\bibfnamefont {H.}~\bibnamefont
  {Gies}}, \bibinfo {author} {\bibfnamefont {C.}~\bibnamefont {Gneiting}}, \
  and\ \bibinfo {author} {\bibfnamefont {R.}~\bibnamefont {Sondenheimer}},\
  }\href {\doibase 10.1103/PhysRevD.89.045012} {\bibfield  {journal} {\bibinfo
  {journal} {Phys. Rev.}\ }\textbf {\bibinfo {volume} {D89}},\ \bibinfo {pages}
  {045012} (\bibinfo {year} {2014})},\ \Eprint {http://arxiv.org/abs/1308.5075}
  {arXiv:1308.5075 [hep-ph]} \BibitemShut {NoStop}%
%%CITATION = ARXIV:1308.5075;%%
\bibitem [{\citenamefont {Wetterich}(1993)}]{Wetterich:1992yh}%
  \BibitemOpen
  \bibfield  {author} {\bibinfo {author} {\bibfnamefont {C.}~\bibnamefont
  {Wetterich}},\ }\href {\doibase 10.1016/0370-2693(93)90726-X} {\bibfield
  {journal} {\bibinfo  {journal} {Phys. Lett.}\ }\textbf {\bibinfo {volume}
  {B301}},\ \bibinfo {pages} {90} (\bibinfo {year} {1993})}\BibitemShut
  {NoStop}%
%%CITATION = PHLTA,B301,90;%%
\bibitem [{\citenamefont {Berges}\ \emph {et~al.}(2002)\citenamefont {Berges},
  \citenamefont {Tetradis},\ and\ \citenamefont {Wetterich}}]{Berges:2000ew}%
  \BibitemOpen
  \bibfield  {author} {\bibinfo {author} {\bibfnamefont {J.}~\bibnamefont
  {Berges}}, \bibinfo {author} {\bibfnamefont {N.}~\bibnamefont {Tetradis}}, \
  and\ \bibinfo {author} {\bibfnamefont {C.}~\bibnamefont {Wetterich}},\ }\href
  {\doibase 10.1016/S0370-1573(01)00098-9} {\bibfield  {journal} {\bibinfo
  {journal} {Phys. Rept.}\ }\textbf {\bibinfo {volume} {363}},\ \bibinfo
  {pages} {223} (\bibinfo {year} {2002})},\ \Eprint
  {http://arxiv.org/abs/hep-ph/0005122} {arXiv:hep-ph/0005122 [hep-ph]}
  \BibitemShut {NoStop}%
%%CITATION = HEP-PH/0005122;%%
\bibitem [{\citenamefont {Pawlowski}(2007)}]{Pawlowski:2005xe}%
  \BibitemOpen
  \bibfield  {author} {\bibinfo {author} {\bibfnamefont {J.~M.}\ \bibnamefont
  {Pawlowski}},\ }\href {\doibase 10.1016/j.aop.2007.01.007} {\bibfield
  {journal} {\bibinfo  {journal} {Annals Phys.}\ }\textbf {\bibinfo {volume}
  {322}},\ \bibinfo {pages} {2831} (\bibinfo {year} {2007})},\ \Eprint
  {http://arxiv.org/abs/hep-th/0512261} {arXiv:hep-th/0512261 [hep-th]}
  \BibitemShut {NoStop}%
%%CITATION = HEP-TH/0512261;%%
\bibitem [{\citenamefont {Gies}(2012)}]{Gies:2006wv}%
  \BibitemOpen
  \bibfield  {author} {\bibinfo {author} {\bibfnamefont {H.}~\bibnamefont
  {Gies}},\ }\bibfield  {booktitle} {\emph {\bibinfo {booktitle} {{ECT* School
  on Renormalization Group and Effective Field Theory Approaches to Many-Body
  Systems Trento, Italy, February 27-March 10, 2006}}},\ }\href {\doibase
  10.1007/978-3-642-27320-9_6} {\bibfield  {journal} {\bibinfo  {journal}
  {Lect. Notes Phys.}\ }\textbf {\bibinfo {volume} {852}},\ \bibinfo {pages}
  {287} (\bibinfo {year} {2012})},\ \Eprint
  {http://arxiv.org/abs/hep-ph/0611146} {arXiv:hep-ph/0611146 [hep-ph]}
  \BibitemShut {NoStop}%
%%CITATION = HEP-PH/0611146;%%
\bibitem [{\citenamefont {Delamotte}(2012)}]{Delamotte:2007pf}%
  \BibitemOpen
  \bibfield  {author} {\bibinfo {author} {\bibfnamefont {B.}~\bibnamefont
  {Delamotte}},\ }\href {\doibase 10.1007/978-3-642-27320-9_2} {\bibfield
  {journal} {\bibinfo  {journal} {Lect. Notes Phys.}\ }\textbf {\bibinfo
  {volume} {852}},\ \bibinfo {pages} {49} (\bibinfo {year} {2012})},\ \Eprint
  {http://arxiv.org/abs/cond-mat/0702365} {arXiv:cond-mat/0702365
  [cond-mat.stat-mech]} \BibitemShut {NoStop}%
%%CITATION = COND-MAT/0702365;%%
\bibitem [{\citenamefont {Braun}(2012)}]{Braun:2011pp}%
  \BibitemOpen
  \bibfield  {author} {\bibinfo {author} {\bibfnamefont {J.}~\bibnamefont
  {Braun}},\ }\href {\doibase 10.1088/0954-3899/39/3/033001} {\bibfield
  {journal} {\bibinfo  {journal} {J. Phys.}\ }\textbf {\bibinfo {volume}
  {G39}},\ \bibinfo {pages} {033001} (\bibinfo {year} {2012})},\ \Eprint
  {http://arxiv.org/abs/1108.4449} {arXiv:1108.4449 [hep-ph]} \BibitemShut
  {NoStop}%
%%CITATION = ARXIV:1108.4449;%%
\bibitem [{\citenamefont {Dupuis}\ \emph {et~al.}(2021)\citenamefont {Dupuis},
  \citenamefont {Canet}, \citenamefont {Eichhorn}, \citenamefont {Metzner},
  \citenamefont {Pawlowski}, \citenamefont {Tissier},\ and\ \citenamefont
  {Wschebor}}]{Dupuis:2020fhh}%
  \BibitemOpen
  \bibfield  {author} {\bibinfo {author} {\bibfnamefont {N.}~\bibnamefont
  {Dupuis}}, \bibinfo {author} {\bibfnamefont {L.}~\bibnamefont {Canet}},
  \bibinfo {author} {\bibfnamefont {A.}~\bibnamefont {Eichhorn}}, \bibinfo
  {author} {\bibfnamefont {W.}~\bibnamefont {Metzner}}, \bibinfo {author}
  {\bibfnamefont {J.~M.}\ \bibnamefont {Pawlowski}}, \bibinfo {author}
  {\bibfnamefont {M.}~\bibnamefont {Tissier}}, \ and\ \bibinfo {author}
  {\bibfnamefont {N.}~\bibnamefont {Wschebor}},\ }\href {\doibase
  10.1016/j.physrep.2021.01.001} {\bibfield  {journal} {\bibinfo  {journal}
  {Phys. Rept.}\ }\textbf {\bibinfo {volume} {910}},\ \bibinfo {pages} {1}
  (\bibinfo {year} {2021})},\ \Eprint {http://arxiv.org/abs/2006.04853}
  {arXiv:2006.04853 [cond-mat.stat-mech]} \BibitemShut {NoStop}%
\bibitem [{\citenamefont {Jungnickel}\ and\ \citenamefont
  {Wetterich}(1996)}]{Jungnickel:1995fp}%
  \BibitemOpen
  \bibfield  {author} {\bibinfo {author} {\bibfnamefont {D.~U.}\ \bibnamefont
  {Jungnickel}}\ and\ \bibinfo {author} {\bibfnamefont {C.}~\bibnamefont
  {Wetterich}},\ }\href {\doibase 10.1103/PhysRevD.53.5142} {\bibfield
  {journal} {\bibinfo  {journal} {Phys. Rev.}\ }\textbf {\bibinfo {volume}
  {D53}},\ \bibinfo {pages} {5142} (\bibinfo {year} {1996})},\ \Eprint
  {http://arxiv.org/abs/hep-ph/9505267} {arXiv:hep-ph/9505267 [hep-ph]}
  \BibitemShut {NoStop}%
%%CITATION = HEP-PH/9505267;%%
\bibitem [{\citenamefont {Hofling}\ \emph {et~al.}(2002)\citenamefont
  {Hofling}, \citenamefont {Nowak},\ and\ \citenamefont
  {Wetterich}}]{Hofling:2002hj}%
  \BibitemOpen
  \bibfield  {author} {\bibinfo {author} {\bibfnamefont {F.}~\bibnamefont
  {Hofling}}, \bibinfo {author} {\bibfnamefont {C.}~\bibnamefont {Nowak}}, \
  and\ \bibinfo {author} {\bibfnamefont {C.}~\bibnamefont {Wetterich}},\ }\href
  {\doibase 10.1103/PhysRevB.66.205111} {\bibfield  {journal} {\bibinfo
  {journal} {Phys. Rev.}\ }\textbf {\bibinfo {volume} {B66}},\ \bibinfo {pages}
  {205111} (\bibinfo {year} {2002})},\ \Eprint
  {http://arxiv.org/abs/cond-mat/0203588} {arXiv:cond-mat/0203588 [cond-mat]}
  \BibitemShut {NoStop}%
%%CITATION = COND-MAT/0203588;%%
\bibitem [{\citenamefont {Diehl}\ \emph {et~al.}(2010)\citenamefont {Diehl},
  \citenamefont {Floerchinger}, \citenamefont {Gies}, \citenamefont
  {Pawlowski},\ and\ \citenamefont {Wetterich}}]{Diehl:2009ma}%
  \BibitemOpen
  \bibfield  {author} {\bibinfo {author} {\bibfnamefont {S.}~\bibnamefont
  {Diehl}}, \bibinfo {author} {\bibfnamefont {S.}~\bibnamefont {Floerchinger}},
  \bibinfo {author} {\bibfnamefont {H.}~\bibnamefont {Gies}}, \bibinfo {author}
  {\bibfnamefont {J.~M.}\ \bibnamefont {Pawlowski}}, \ and\ \bibinfo {author}
  {\bibfnamefont {C.}~\bibnamefont {Wetterich}},\ }\href {\doibase
  10.1002/andp.201010458} {\bibfield  {journal} {\bibinfo  {journal} {Annalen
  Phys.}\ }\textbf {\bibinfo {volume} {522}},\ \bibinfo {pages} {615} (\bibinfo
  {year} {2010})},\ \Eprint {http://arxiv.org/abs/0907.2193} {arXiv:0907.2193
  [cond-mat.quant-gas]} \BibitemShut {NoStop}%
%%CITATION = ARXIV:0907.2193;%%
\bibitem [{\citenamefont {Braun}\ \emph {et~al.}(2011)\citenamefont {Braun},
  \citenamefont {Gies},\ and\ \citenamefont {Scherer}}]{Braun:2010tt}%
  \BibitemOpen
  \bibfield  {author} {\bibinfo {author} {\bibfnamefont {J.}~\bibnamefont
  {Braun}}, \bibinfo {author} {\bibfnamefont {H.}~\bibnamefont {Gies}}, \ and\
  \bibinfo {author} {\bibfnamefont {D.~D.}\ \bibnamefont {Scherer}},\ }\href
  {\doibase 10.1103/PhysRevD.83.085012} {\bibfield  {journal} {\bibinfo
  {journal} {Phys. Rev.}\ }\textbf {\bibinfo {volume} {D83}},\ \bibinfo {pages}
  {085012} (\bibinfo {year} {2011})},\ \Eprint {http://arxiv.org/abs/1011.1456}
  {arXiv:1011.1456 [hep-th]} \BibitemShut {NoStop}%
%%CITATION = ARXIV:1011.1456;%%
\bibitem [{\citenamefont {Mesterhazy}\ \emph {et~al.}(2012)\citenamefont
  {Mesterhazy}, \citenamefont {Berges},\ and\ \citenamefont {von
  Smekal}}]{Mesterhazy:2012ei}%
  \BibitemOpen
  \bibfield  {author} {\bibinfo {author} {\bibfnamefont {D.}~\bibnamefont
  {Mesterhazy}}, \bibinfo {author} {\bibfnamefont {J.}~\bibnamefont {Berges}},
  \ and\ \bibinfo {author} {\bibfnamefont {L.}~\bibnamefont {von Smekal}},\
  }\href {\doibase 10.1103/PhysRevB.86.245431} {\bibfield  {journal} {\bibinfo
  {journal} {Phys. Rev.}\ }\textbf {\bibinfo {volume} {B86}},\ \bibinfo {pages}
  {245431} (\bibinfo {year} {2012})},\ \Eprint {http://arxiv.org/abs/1207.4054}
  {arXiv:1207.4054 [cond-mat.str-el]} \BibitemShut {NoStop}%
%%CITATION = ARXIV:1207.4054;%%
\bibitem [{\citenamefont {Jakov\'{a}c}\ \emph {et~al.}(2015)\citenamefont
  {Jakov\'{a}c}, \citenamefont {Patk\'{o}s},\ and\ \citenamefont
  {P\'{o}sfay}}]{Jakovac:2014lqa}%
  \BibitemOpen
  \bibfield  {author} {\bibinfo {author} {\bibfnamefont {A.}~\bibnamefont
  {Jakov\'{a}c}}, \bibinfo {author} {\bibfnamefont {A.}~\bibnamefont
  {Patk\'{o}s}}, \ and\ \bibinfo {author} {\bibfnamefont {P.}~\bibnamefont
  {P\'{o}sfay}},\ }\href {\doibase 10.1140/epjc/s10052-014-3228-1} {\bibfield
  {journal} {\bibinfo  {journal} {Eur. Phys. J.}\ }\textbf {\bibinfo {volume}
  {C75}},\ \bibinfo {pages} {2} (\bibinfo {year} {2015})},\ \Eprint
  {http://arxiv.org/abs/1406.3195} {arXiv:1406.3195 [hep-th]} \BibitemShut
  {NoStop}%
%%CITATION = ARXIV:1406.3195;%%
\bibitem [{\citenamefont {Janssen}\ and\ \citenamefont
  {Herbut}(2014)}]{Janssen:2014gea}%
  \BibitemOpen
  \bibfield  {author} {\bibinfo {author} {\bibfnamefont {L.}~\bibnamefont
  {Janssen}}\ and\ \bibinfo {author} {\bibfnamefont {I.~F.}\ \bibnamefont
  {Herbut}},\ }\href {\doibase 10.1103/PhysRevB.89.205403} {\bibfield
  {journal} {\bibinfo  {journal} {Phys. Rev.}\ }\textbf {\bibinfo {volume}
  {B89}},\ \bibinfo {pages} {205403} (\bibinfo {year} {2014})},\ \Eprint
  {http://arxiv.org/abs/1402.6277} {arXiv:1402.6277 [cond-mat.str-el]}
  \BibitemShut {NoStop}%
%%CITATION = ARXIV:1402.6277;%%
\bibitem [{\citenamefont {Vacca}\ and\ \citenamefont
  {Zambelli}(2015)}]{Vacca:2015nta}%
  \BibitemOpen
  \bibfield  {author} {\bibinfo {author} {\bibfnamefont {G.~P.}\ \bibnamefont
  {Vacca}}\ and\ \bibinfo {author} {\bibfnamefont {L.}~\bibnamefont
  {Zambelli}},\ }\href {\doibase 10.1103/PhysRevD.91.125003} {\bibfield
  {journal} {\bibinfo  {journal} {Phys. Rev.}\ }\textbf {\bibinfo {volume}
  {D91}},\ \bibinfo {pages} {125003} (\bibinfo {year} {2015})},\ \Eprint
  {http://arxiv.org/abs/1503.09136} {arXiv:1503.09136 [hep-th]} \BibitemShut
  {NoStop}%
%%CITATION = ARXIV:1503.09136;%%
\bibitem [{\citenamefont {Gies}\ and\ \citenamefont
  {Sondenheimer}(2015)}]{Gies:2014xha}%
  \BibitemOpen
  \bibfield  {author} {\bibinfo {author} {\bibfnamefont {H.}~\bibnamefont
  {Gies}}\ and\ \bibinfo {author} {\bibfnamefont {R.}~\bibnamefont
  {Sondenheimer}},\ }\href {\doibase 10.1140/epjc/s10052-015-3284-1} {\bibfield
   {journal} {\bibinfo  {journal} {Eur. Phys. J.}\ }\textbf {\bibinfo {volume}
  {C75}},\ \bibinfo {pages} {68} (\bibinfo {year} {2015})},\ \Eprint
  {http://arxiv.org/abs/1407.8124} {arXiv:1407.8124 [hep-ph]} \BibitemShut
  {NoStop}%
%%CITATION = ARXIV:1407.8124;%%
\bibitem [{\citenamefont {Classen}\ \emph {et~al.}(2016)\citenamefont
  {Classen}, \citenamefont {Herbut}, \citenamefont {Janssen},\ and\
  \citenamefont {Scherer}}]{Classen:2015mar}%
  \BibitemOpen
  \bibfield  {author} {\bibinfo {author} {\bibfnamefont {L.}~\bibnamefont
  {Classen}}, \bibinfo {author} {\bibfnamefont {I.~F.}\ \bibnamefont {Herbut}},
  \bibinfo {author} {\bibfnamefont {L.}~\bibnamefont {Janssen}}, \ and\
  \bibinfo {author} {\bibfnamefont {M.~M.}\ \bibnamefont {Scherer}},\ }\href
  {\doibase 10.1103/PhysRevB.93.125119} {\bibfield  {journal} {\bibinfo
  {journal} {Phys. Rev.}\ }\textbf {\bibinfo {volume} {B93}},\ \bibinfo {pages}
  {125119} (\bibinfo {year} {2016})},\ \Eprint
  {http://arxiv.org/abs/1510.09003} {arXiv:1510.09003 [cond-mat.str-el]}
  \BibitemShut {NoStop}%
%%CITATION = ARXIV:1510.09003;%%
\bibitem [{\citenamefont {Knorr}(2016)}]{Knorr:2016sfs}%
  \BibitemOpen
  \bibfield  {author} {\bibinfo {author} {\bibfnamefont {B.}~\bibnamefont
  {Knorr}},\ }\href {\doibase 10.1103/PhysRevB.94.245102} {\bibfield  {journal}
  {\bibinfo  {journal} {Phys. Rev.}\ }\textbf {\bibinfo {volume} {B94}},\
  \bibinfo {pages} {245102} (\bibinfo {year} {2016})},\ \Eprint
  {http://arxiv.org/abs/1609.03824} {arXiv:1609.03824 [cond-mat.str-el]}
  \BibitemShut {NoStop}%
%%CITATION = ARXIV:1609.03824;%%
\bibitem [{\citenamefont {Fu}\ \emph {et~al.}(2016)\citenamefont {Fu},
  \citenamefont {Pawlowski}, \citenamefont {Rennecke},\ and\ \citenamefont
  {Schaefer}}]{Fu:2016tey}%
  \BibitemOpen
  \bibfield  {author} {\bibinfo {author} {\bibfnamefont {W.-j.}\ \bibnamefont
  {Fu}}, \bibinfo {author} {\bibfnamefont {J.~M.}\ \bibnamefont {Pawlowski}},
  \bibinfo {author} {\bibfnamefont {F.}~\bibnamefont {Rennecke}}, \ and\
  \bibinfo {author} {\bibfnamefont {B.-J.}\ \bibnamefont {Schaefer}},\ }\href
  {\doibase 10.1103/PhysRevD.94.116020} {\bibfield  {journal} {\bibinfo
  {journal} {Phys. Rev. D}\ }\textbf {\bibinfo {volume} {94}},\ \bibinfo
  {pages} {116020} (\bibinfo {year} {2016})},\ \Eprint
  {http://arxiv.org/abs/1608.04302} {arXiv:1608.04302 [hep-ph]} \BibitemShut
  {NoStop}%
\bibitem [{\citenamefont {Stoll}\ \emph {et~al.}(2021)\citenamefont {Stoll},
  \citenamefont {Zorbach}, \citenamefont {Koenigstein}, \citenamefont {Steil},\
  and\ \citenamefont {Rechenberger}}]{Stoll:2021ori}%
  \BibitemOpen
  \bibfield  {author} {\bibinfo {author} {\bibfnamefont {J.}~\bibnamefont
  {Stoll}}, \bibinfo {author} {\bibfnamefont {N.}~\bibnamefont {Zorbach}},
  \bibinfo {author} {\bibfnamefont {A.}~\bibnamefont {Koenigstein}}, \bibinfo
  {author} {\bibfnamefont {M.~J.}\ \bibnamefont {Steil}}, \ and\ \bibinfo
  {author} {\bibfnamefont {S.}~\bibnamefont {Rechenberger}},\ }\href@noop {} {\
   (\bibinfo {year} {2021})},\ \Eprint {http://arxiv.org/abs/2108.10616}
  {arXiv:2108.10616 [hep-ph]} \BibitemShut {NoStop}%
\bibitem [{\citenamefont {Gies}\ \emph {et~al.}(2025)\citenamefont {Gies},
  \citenamefont {Schmieden},\ and\ \citenamefont {Zambelli}}]{Gies:2023jzd}%
  \BibitemOpen
  \bibfield  {author} {\bibinfo {author} {\bibfnamefont {H.}~\bibnamefont
  {Gies}}, \bibinfo {author} {\bibfnamefont {R.}~\bibnamefont {Schmieden}}, \
  and\ \bibinfo {author} {\bibfnamefont {L.}~\bibnamefont {Zambelli}},\ }\href
  {\doibase 10.1140/epjc/s10052-024-13689-3} {\bibfield  {journal} {\bibinfo
  {journal} {Eur. Phys. J. C}\ }\textbf {\bibinfo {volume} {85}},\ \bibinfo
  {pages} {56} (\bibinfo {year} {2025})},\ \Eprint
  {http://arxiv.org/abs/2306.05943} {arXiv:2306.05943 [hep-th]} \BibitemShut
  {NoStop}%
\bibitem [{\citenamefont {Bervillier}\ \emph {et~al.}(2007)\citenamefont
  {Bervillier}, \citenamefont {Juttner},\ and\ \citenamefont
  {Litim}}]{Bervillier:2007rc}%
  \BibitemOpen
  \bibfield  {author} {\bibinfo {author} {\bibfnamefont {C.}~\bibnamefont
  {Bervillier}}, \bibinfo {author} {\bibfnamefont {A.}~\bibnamefont {Juttner}},
  \ and\ \bibinfo {author} {\bibfnamefont {D.~F.}\ \bibnamefont {Litim}},\
  }\href {\doibase 10.1016/j.nuclphysb.2007.03.036} {\bibfield  {journal}
  {\bibinfo  {journal} {Nucl. Phys. B}\ }\textbf {\bibinfo {volume} {783}},\
  \bibinfo {pages} {213} (\bibinfo {year} {2007})},\ \Eprint
  {http://arxiv.org/abs/hep-th/0701172} {arXiv:hep-th/0701172} \BibitemShut
  {NoStop}%
\bibitem [{\citenamefont {Borchardt}\ and\ \citenamefont
  {Knorr}(2015)}]{Borchardt:2015rxa}%
  \BibitemOpen
  \bibfield  {author} {\bibinfo {author} {\bibfnamefont {J.}~\bibnamefont
  {Borchardt}}\ and\ \bibinfo {author} {\bibfnamefont {B.}~\bibnamefont
  {Knorr}},\ }\href {\doibase 10.1103/PhysRevD.91.105011} {\bibfield  {journal}
  {\bibinfo  {journal} {Phys. Rev.}\ }\textbf {\bibinfo {volume} {D91}},\
  \bibinfo {pages} {105011} (\bibinfo {year} {2015})},\ \Eprint
  {http://arxiv.org/abs/1502.07511} {arXiv:1502.07511 [hep-th]} \BibitemShut
  {NoStop}%
%%CITATION = ARXIV:1502.07511;%%
\bibitem [{\citenamefont {Borchardt}\ and\ \citenamefont
  {Knorr}(2016)}]{Borchardt:2016pif}%
  \BibitemOpen
  \bibfield  {author} {\bibinfo {author} {\bibfnamefont {J.}~\bibnamefont
  {Borchardt}}\ and\ \bibinfo {author} {\bibfnamefont {B.}~\bibnamefont
  {Knorr}},\ }\href {\doibase 10.1103/PhysRevD.94.025027} {\bibfield  {journal}
  {\bibinfo  {journal} {Phys. Rev.}\ }\textbf {\bibinfo {volume} {D94}},\
  \bibinfo {pages} {025027} (\bibinfo {year} {2016})},\ \Eprint
  {http://arxiv.org/abs/1603.06726} {arXiv:1603.06726 [hep-th]} \BibitemShut
  {NoStop}%
%%CITATION = ARXIV:1603.06726;%%
\bibitem [{\citenamefont {Borchardt}\ \emph {et~al.}(2016)\citenamefont
  {Borchardt}, \citenamefont {Gies},\ and\ \citenamefont
  {Sondenheimer}}]{Borchardt:2016xju}%
  \BibitemOpen
  \bibfield  {author} {\bibinfo {author} {\bibfnamefont {J.}~\bibnamefont
  {Borchardt}}, \bibinfo {author} {\bibfnamefont {H.}~\bibnamefont {Gies}}, \
  and\ \bibinfo {author} {\bibfnamefont {R.}~\bibnamefont {Sondenheimer}},\
  }\href {\doibase 10.1140/epjc/s10052-016-4300-9} {\bibfield  {journal}
  {\bibinfo  {journal} {Eur. Phys. J.}\ }\textbf {\bibinfo {volume} {C76}},\
  \bibinfo {pages} {472} (\bibinfo {year} {2016})},\ \Eprint
  {http://arxiv.org/abs/1603.05861} {arXiv:1603.05861 [hep-ph]} \BibitemShut
  {NoStop}%
%%CITATION = ARXIV:1603.05861;%%
\bibitem [{\citenamefont {Grossi}\ and\ \citenamefont
  {Wink}(2023)}]{Grossi:2019urj}%
  \BibitemOpen
  \bibfield  {author} {\bibinfo {author} {\bibfnamefont {E.}~\bibnamefont
  {Grossi}}\ and\ \bibinfo {author} {\bibfnamefont {N.}~\bibnamefont {Wink}},\
  }\href {\doibase 10.21468/SciPostPhysCore.6.4.071} {\bibfield  {journal}
  {\bibinfo  {journal} {SciPost Phys. Core}\ }\textbf {\bibinfo {volume} {6}},\
  \bibinfo {pages} {071} (\bibinfo {year} {2023})},\ \Eprint
  {http://arxiv.org/abs/1903.09503} {arXiv:1903.09503 [hep-th]} \BibitemShut
  {NoStop}%
\bibitem [{\citenamefont {Koenigstein}\ \emph {et~al.}(2022)\citenamefont
  {Koenigstein}, \citenamefont {Steil}, \citenamefont {Wink}, \citenamefont
  {Grossi}, \citenamefont {Braun}, \citenamefont {Buballa},\ and\ \citenamefont
  {Rischke}}]{Koenigstein:2021syz}%
  \BibitemOpen
  \bibfield  {author} {\bibinfo {author} {\bibfnamefont {A.}~\bibnamefont
  {Koenigstein}}, \bibinfo {author} {\bibfnamefont {M.~J.}\ \bibnamefont
  {Steil}}, \bibinfo {author} {\bibfnamefont {N.}~\bibnamefont {Wink}},
  \bibinfo {author} {\bibfnamefont {E.}~\bibnamefont {Grossi}}, \bibinfo
  {author} {\bibfnamefont {J.}~\bibnamefont {Braun}}, \bibinfo {author}
  {\bibfnamefont {M.}~\bibnamefont {Buballa}}, \ and\ \bibinfo {author}
  {\bibfnamefont {D.~H.}\ \bibnamefont {Rischke}},\ }\href {\doibase
  10.1103/PhysRevD.106.065012} {\bibfield  {journal} {\bibinfo  {journal}
  {Phys. Rev. D}\ }\textbf {\bibinfo {volume} {106}},\ \bibinfo {pages}
  {065012} (\bibinfo {year} {2022})},\ \Eprint
  {http://arxiv.org/abs/2108.02504} {arXiv:2108.02504 [cond-mat.stat-mech]}
  \BibitemShut {NoStop}%
\bibitem [{\citenamefont {Ihssen}\ \emph {et~al.}(2024)\citenamefont {Ihssen},
  \citenamefont {Pawlowski}, \citenamefont {Sattler},\ and\ \citenamefont
  {Wink}}]{Ihssen:2022xkr}%
  \BibitemOpen
  \bibfield  {author} {\bibinfo {author} {\bibfnamefont {F.}~\bibnamefont
  {Ihssen}}, \bibinfo {author} {\bibfnamefont {J.~M.}\ \bibnamefont
  {Pawlowski}}, \bibinfo {author} {\bibfnamefont {F.~R.}\ \bibnamefont
  {Sattler}}, \ and\ \bibinfo {author} {\bibfnamefont {N.}~\bibnamefont
  {Wink}},\ }\href {\doibase 10.1016/j.cpc.2024.109182} {\bibfield  {journal}
  {\bibinfo  {journal} {Comput. Phys. Commun.}\ }\textbf {\bibinfo {volume}
  {300}},\ \bibinfo {pages} {109182} (\bibinfo {year} {2024})},\ \Eprint
  {http://arxiv.org/abs/2207.12266} {arXiv:2207.12266 [hep-th]} \BibitemShut
  {NoStop}%
\bibitem [{\citenamefont {Ihssen}\ \emph {et~al.}(2023)\citenamefont {Ihssen},
  \citenamefont {Sattler},\ and\ \citenamefont {Wink}}]{Ihssen:2023qaq}%
  \BibitemOpen
  \bibfield  {author} {\bibinfo {author} {\bibfnamefont {F.}~\bibnamefont
  {Ihssen}}, \bibinfo {author} {\bibfnamefont {F.~R.}\ \bibnamefont {Sattler}},
  \ and\ \bibinfo {author} {\bibfnamefont {N.}~\bibnamefont {Wink}},\ }\href
  {\doibase 10.1103/PhysRevD.107.114009} {\bibfield  {journal} {\bibinfo
  {journal} {Phys. Rev. D}\ }\textbf {\bibinfo {volume} {107}},\ \bibinfo
  {pages} {114009} (\bibinfo {year} {2023})},\ \Eprint
  {http://arxiv.org/abs/2302.04736} {arXiv:2302.04736 [hep-th]} \BibitemShut
  {NoStop}%
\bibitem [{\citenamefont {Sattler}\ and\ \citenamefont
  {Pawlowski}(2024)}]{Sattler:2024ozv}%
  \BibitemOpen
  \bibfield  {author} {\bibinfo {author} {\bibfnamefont {F.~R.}\ \bibnamefont
  {Sattler}}\ and\ \bibinfo {author} {\bibfnamefont {J.~M.}\ \bibnamefont
  {Pawlowski}},\ }\href@noop {} {\  (\bibinfo {year} {2024})},\ \Eprint
  {http://arxiv.org/abs/2412.13043} {arXiv:2412.13043 [hep-ph]} \BibitemShut
  {NoStop}%
\bibitem [{\citenamefont {O'Raifeartaigh}\ \emph {et~al.}(1986)\citenamefont
  {O'Raifeartaigh}, \citenamefont {Wipf},\ and\ \citenamefont
  {Yoneyama}}]{ORaifeartaigh:1986hi}%
  \BibitemOpen
  \bibfield  {author} {\bibinfo {author} {\bibfnamefont {L.}~\bibnamefont
  {O'Raifeartaigh}}, \bibinfo {author} {\bibfnamefont {A.}~\bibnamefont
  {Wipf}}, \ and\ \bibinfo {author} {\bibfnamefont {H.}~\bibnamefont
  {Yoneyama}},\ }\href {\doibase 10.1016/S0550-3213(86)80031-1} {\bibfield
  {journal} {\bibinfo  {journal} {Nucl. Phys.}\ }\textbf {\bibinfo {volume}
  {B271}},\ \bibinfo {pages} {653} (\bibinfo {year} {1986})}\BibitemShut
  {NoStop}%
%%CITATION = NUPHA,B271,653;%%
\bibitem [{\citenamefont {Litim}\ \emph {et~al.}(2006)\citenamefont {Litim},
  \citenamefont {Pawlowski},\ and\ \citenamefont {Vergara}}]{Litim:2006nn}%
  \BibitemOpen
  \bibfield  {author} {\bibinfo {author} {\bibfnamefont {D.~F.}\ \bibnamefont
  {Litim}}, \bibinfo {author} {\bibfnamefont {J.~M.}\ \bibnamefont
  {Pawlowski}}, \ and\ \bibinfo {author} {\bibfnamefont {L.}~\bibnamefont
  {Vergara}},\ }\href
  {https://inspirehep.net/record/710544/files/arXiv:hep-th_0602140.pdf?version=2}
  {\  (\bibinfo {year} {2006})},\ \Eprint {http://arxiv.org/abs/hep-th/0602140}
  {arXiv:hep-th/0602140 [hep-th]} \BibitemShut {NoStop}%
%%CITATION = HEP-TH/0602140;%%
\bibitem [{\citenamefont {Zorbach}\ \emph
  {et~al.}(2024{\natexlab{a}})\citenamefont {Zorbach}, \citenamefont {Stoll},\
  and\ \citenamefont {Braun}}]{Zorbach:2024zjx}%
  \BibitemOpen
  \bibfield  {author} {\bibinfo {author} {\bibfnamefont {N.}~\bibnamefont
  {Zorbach}}, \bibinfo {author} {\bibfnamefont {J.}~\bibnamefont {Stoll}}, \
  and\ \bibinfo {author} {\bibfnamefont {J.}~\bibnamefont {Braun}},\
  }\href@noop {} {\  (\bibinfo {year} {2024}{\natexlab{a}})},\ \Eprint
  {http://arxiv.org/abs/2401.12854} {arXiv:2401.12854 [hep-ph]} \BibitemShut
  {NoStop}%
\bibitem [{\citenamefont {Zorbach}\ \emph
  {et~al.}(2024{\natexlab{b}})\citenamefont {Zorbach}, \citenamefont
  {Koenigstein},\ and\ \citenamefont {Braun}}]{Zorbach:2024rre}%
  \BibitemOpen
  \bibfield  {author} {\bibinfo {author} {\bibfnamefont {N.}~\bibnamefont
  {Zorbach}}, \bibinfo {author} {\bibfnamefont {A.}~\bibnamefont
  {Koenigstein}}, \ and\ \bibinfo {author} {\bibfnamefont {J.}~\bibnamefont
  {Braun}},\ }\href@noop {} {\  (\bibinfo {year} {2024}{\natexlab{b}})},\
  \Eprint {http://arxiv.org/abs/2412.16053} {arXiv:2412.16053
  [cond-mat.stat-mech]} \BibitemShut {NoStop}%
\bibitem [{\citenamefont {Gies}\ \emph {et~al.}(2013)\citenamefont {Gies},
  \citenamefont {Rechenberger}, \citenamefont {Scherer},\ and\ \citenamefont
  {Zambelli}}]{Gies:2013pma}%
  \BibitemOpen
  \bibfield  {author} {\bibinfo {author} {\bibfnamefont {H.}~\bibnamefont
  {Gies}}, \bibinfo {author} {\bibfnamefont {S.}~\bibnamefont {Rechenberger}},
  \bibinfo {author} {\bibfnamefont {M.~M.}\ \bibnamefont {Scherer}}, \ and\
  \bibinfo {author} {\bibfnamefont {L.}~\bibnamefont {Zambelli}},\ }\href
  {\doibase 10.1140/epjc/s10052-013-2652-y} {\bibfield  {journal} {\bibinfo
  {journal} {Eur. Phys. J.}\ }\textbf {\bibinfo {volume} {C73}},\ \bibinfo
  {pages} {2652} (\bibinfo {year} {2013})},\ \Eprint
  {http://arxiv.org/abs/1306.6508} {arXiv:1306.6508 [hep-th]} \BibitemShut
  {NoStop}%
%%CITATION = ARXIV:1306.6508;%%
\bibitem [{\citenamefont {Gies}\ and\ \citenamefont
  {Ziebell}(2020)}]{Gies:2020xuh}%
  \BibitemOpen
  \bibfield  {author} {\bibinfo {author} {\bibfnamefont {H.}~\bibnamefont
  {Gies}}\ and\ \bibinfo {author} {\bibfnamefont {J.}~\bibnamefont {Ziebell}},\
  }\href {\doibase 10.1140/epjc/s10052-020-8171-8} {\bibfield  {journal}
  {\bibinfo  {journal} {Eur. Phys. J. C}\ }\textbf {\bibinfo {volume} {80}},\
  \bibinfo {pages} {607} (\bibinfo {year} {2020})},\ \Eprint
  {http://arxiv.org/abs/2005.07586} {arXiv:2005.07586 [hep-th]} \BibitemShut
  {NoStop}%
\bibitem [{\citenamefont {Litim}(2000)}]{Litim:2000ci}%
  \BibitemOpen
  \bibfield  {author} {\bibinfo {author} {\bibfnamefont {D.~F.}\ \bibnamefont
  {Litim}},\ }\href {\doibase 10.1016/S0370-2693(00)00748-6} {\bibfield
  {journal} {\bibinfo  {journal} {Phys. Lett.}\ }\textbf {\bibinfo {volume}
  {B486}},\ \bibinfo {pages} {92} (\bibinfo {year} {2000})},\ \Eprint
  {http://arxiv.org/abs/hep-th/0005245} {arXiv:hep-th/0005245 [hep-th]}
  \BibitemShut {NoStop}%
%%CITATION = HEP-TH/0005245;%%
\bibitem [{\citenamefont {Litim}(2001)}]{Litim:2001up}%
  \BibitemOpen
  \bibfield  {author} {\bibinfo {author} {\bibfnamefont {D.~F.}\ \bibnamefont
  {Litim}},\ }\href {\doibase 10.1103/PhysRevD.64.105007} {\bibfield  {journal}
  {\bibinfo  {journal} {Phys. Rev.}\ }\textbf {\bibinfo {volume} {D64}},\
  \bibinfo {pages} {105007} (\bibinfo {year} {2001})},\ \Eprint
  {http://arxiv.org/abs/hep-th/0103195} {arXiv:hep-th/0103195 [hep-th]}
  \BibitemShut {NoStop}%
%%CITATION = HEP-TH/0103195;%%
\end{thebibliography}%

% \newpage
% \newpage
% \clearpage
%\onecolumngrid

\end{document}